\documentclass[aps,nofootinbib,superscriptaddress]{revtex4}

\usepackage{amsmath}
\usepackage{graphicx}
\usepackage{float}
\usepackage{bm}
\usepackage{url}

\begin{document}

\title{$\pi-N$ Drell-Yan process in TMD factorization}
\author{Xiaoyu Wang}
\affiliation{School of Physics and Engineering, Zhengzhou University, Zhengzhou, Henan 450001, China}
\author{Zhun Lu}
\email{zhunlu@seu.edu.cn}
\affiliation{School of Physics, Southeast University, Nanjing 211189, China}

\begin{abstract}
This article presents the review of the current understanding on the pion-nucleon Drell-Yan process from the point of view of the TMD factorization.
Using the evolution formalism for the unpolarized and polarized TMD distributions developed recently, we provide the theoretical expression of the relevant physical observables, namely, the unpolarized cross section, the Sivers asymmetry, and the $\cos2\phi$ asymmetry contributed by the double Boer-Mulders effects.
The corresponding phenomenology, particularly at the kinematical configuration of the COMPASS $\pi N$ Drell-Yan facility, is displayed numerically.
\end{abstract}

\maketitle

\section{Introduction}

After the first observation of the $\mu^+ \mu^-$ lepton pairs produced in $p\,N$ collisions~\cite{Christenson:1970um}, the process was interpreted that a quark and an antiquark from each initial hadron annihilate into a virtual photon, which in turn decays into a lepton pair~\cite{Drell:1970wh}. This explanation makes the process an ideal tool to explore the internal structure of both the beam and target hadrons. Since then, a wide range of studies on this (Drell-Yan) process have been carried out.
In particular, the $\pi N$ Drell-Yan process has the unique capability to pin down the partonic structure of the pion, which is an unstable particle and therefore cannot serve as a target in deep inelastic scattering processes.
Several pion induced experiments have been carried out, such as the NA10 experiment at CERN~\cite{Bordalo:1987cs,Betev:1985pf,Falciano:1986wk,Guanziroli:1987rp}, the E615~\cite{Conway:1989fs}, E444~\cite{Palestini:1985zc} and E537~\cite{Anassontzis:1987hk} experiments at Fermilab three decades ago.
These experimental measurements have provided plenty of data, which have been used to considerably constrain the distribution function of the pion meson.
Recently, a new pion-induced Drell-Yan program with polarized target was also proposed~\cite{Gautheron:2010wva} at the COMPASS of CERN, and the first data using a high-intensity $\pi$ beam of 190 GeV colliding on a NH$_3$ target has already come out~\cite{Aghasyan:2017jop}.

Bulk of the events in the Drell-Yan reaction are from the region where the transverse momentum of the dilepton $q_\perp$ is much smaller than the mass $Q$ of the virtual vector boson, thus the intrinsic transverse momenta of initial partons become relevant. It is also the most interesting regime where a lot of intriguing physics arises. Moreover, in the small $q_\perp$ region ($q_\perp \sim \Lambda_{\textrm QCD}$), the fixed-order calculations of the cross sections in the collinear picture fail, leading to large double logarithms of the type $\alpha \ln^2(q_\perp^2/Q^2)$.
It is necessary to resum such logarithmic contributions to all orders in the strong coupling $\alpha_s$ to obtain a reliable result.
The standard approach for such resummation is the Collins-Soper-Sterman (CSS) formalism~\cite{Collins:1984kg}, originated from previous work on the Drell-Yan process and the $e^+ e^-$ annihilation three decades ago. In recent years the CSS formalism has been successfully applied to develop a factorization theorem~\cite{Ji:2004xq,Ji:2004wu,Collins:2011zzd} in which the gauge-invariant~\cite{Collins:2002kn,Ji:2002aa,Belitsky:2002sm,Boer:2003cm} transverse momentum dependent (TMD) parton distribution functions or fragmentation functions (collectively called as TMDs)~\cite{Mulders:1995dh,Bacchetta:2006tn} play a central role.
From the point of view of TMD factorization~\cite{Collins:1981uk,Collins:1984kg,Collins:2011zzd,Ji:2004xq}, physical observables can be written as convolutions of a factor related to hard scattering and well-defined TMDs.
After solving the evolution equations, the TMDs at fixed energy scale can be expressed as a convolution of their collinear counterparts and perturbatively calculable coefficients in the perturbative region,
and the evolution from one energy scale to another energy scale is included in the exponential factor of the so-called Sudakov-like form factors~\cite{Collins:1984kg,Collins:2011zzd,Aybat:2011zv,Collins:1999dz}.
The TMD factorization has been widely applied to various high energy processes, such as the semi-inclusive deep inelastic scatering (SIDIS)~\cite{Collins:1981uk,Collins:2011zzd,Ji:2004wu,Aybat:2011zv,Collins:2012uy,Echevarria:2012pw},
$e^+ e^-$ annihilation~\cite{Collins:2011zzd,Pitonyak:2013dsu,Boer:2008fr}, Drell-Yan~\cite{Collins:2011zzd,Arnold:2008kf} and W/Z production in hadron collision~\cite{Collins:2011zzd,Collins:1984kg,Lambertsen:2016wgj}.
The TMD factorization can be also extended to the moderate $q_\perp$ region where an equivalence~\cite{Ji:2006ub,Ji:2006vf} between the TMD factorization and the twist-3 collinear factorization is found.

One of the most important observables in the polarized Drell-Yan process is the Sivers asymmetry.
It is contributed by the so called Sivers function~\cite{Sivers:1989cc}, a time-reversal-odd (T-odd) distribution describing the asymmetric distribution of unpolarized quarks inside a transversely polarized nucleon through the correlation between the quark transverse momentum and the nucleon transverse spin. Remarkably, QCD predicts that the sign of the Sivers function changes in SIDIS with respect to the Drell-Yan process~\cite{Brodsky:2002cx,Brodsky:2002rv,Collins:2002kn}. The verification of this sign change~\cite{Anselmino:2009st,Kang:2009bp,Peng:2014hta,Echevarria:2014xaa,Huang:2015vpy,Anselmino:2016uie} is one of the most fundamental tests of our understanding of the QCD dynamics and the factorization schemes, and it is also the main pursue of the existing and future Drell-Yan facilities~\cite{Aghasyan:2017jop,Gautheron:2010wva,Fermilab1,Fermilab2,ANDY,Adamczyk:2015gyk}. The advantage of the $\pi\,N$ Drell-Yan measurement at COMPASS is that almost the same setup~\cite{Aghasyan:2017jop,Adolph:2016dvl} is used in SIDIS and Drell-Yan processes, which may reduce the uncertainty in the extraction of the Sivers function.
In particular, the COMPASS collaboration measured for the first time the transverse-spin-dependent azimuthal asymmetries~\cite{Aghasyan:2017jop} in the $\pi^- N$ Drell-Yan process.

Another important observable in the Drell-Yan process is the $\cos 2\phi$ angular asymmetry, where $\phi$ corresponds to the azimuthal angle of the dilepton. The fixed-target measurements from the NA10 and E615 collaborations showed that the unpolarized cross section possesses large $\cos 2\phi$ asymmetry, which violates the Lam-Tung relation~\cite{Lam:1978pu}.
Similar violation has also been observed in the $pp$ colliders at Tevatron~\cite{Aaltonen:2011nr} and LHC~\cite{Khachatryan:2015paa}.
It has been explained from the viewpoints of higher-twist effect~\cite{bran93,bran94,Eskola}, the non-coplanarity effect~\cite{Peng:2015spa,Lambertsen:2016wgj} and the QCD radiative effects at higher order~\cite{Boer:2006eq,Chang:2018pvk}. Another promising origin~\cite{Boer:1999mm} for the violation of the Lam-Tung relation at low transverse momentum is the convolution of the two Boer-Mulders functions~\cite{Boer:1997nt} from each hadron.
The Boer-Mulders function is also a TMD distribution. As the chiral-odd partner of the Sivers function, it describes the transverse-polarization asymmetry of quarks inside an
unpolarized hadron~\cite{Boer:1999mm,Boer:1997nt}, thereby allowing the probe of the transverse spin physics from unpolarized reaction.

This article aims at a review on the current status of our understanding on the Drell-Yan dilepton production at low transverse momentum, especially from the $\pi N$ collision, based on the recent development of the TMD factorization.
We will mainly focus on the phenomenology of the Sivers asymmetry as well as the $\cos 2\phi$ asymmetry from the double Boer-Mulders effect. In order to quantitatively understand various spin/azimuthal asymmetries in the $\pi N$ Drell-Yan process, a particularly important step is to know in high accuracy the  spin-averaged differential cross-section of the same process with azimuthal angles integrated out, since it always appears in the denominator of the asymmetries' definition. Thus, the spin-averaged cross-section will be also discussed in great details.

The remained content of the article is organised as follows. In Sec.~II, we will review the TMD evolution formalism of the TMDs, mostly following the approach established in Ref.~\cite{Collins:2011zzd}. Particularly, we will discuss in details the extraction of the nonperturbative Sudakov form factor for the unpolarized TMD distribution of the proton/pion as well as that for the Sivers function. In Section. III, putting the evolved result of the TMD distributions into the TMD factorization formulae, we will present the theoretical expression of the physical observables, such as the unpolarized differential cross-section, the Sivers asymmetry, and the $\cos2\phi$ asymmetry contributed by the double Boer-Mulders effect. In Section.~IV, we present the numerical evolution results of the unpolarized TMD distributions and the Boer-Mulders function of the pion meson, as well as that of the Sivers function of the proton. In Section.~V, we display the phenomenology of the physical observables (unpolarized differential cross-section, the Sivers asymmetry and the $\cos2\phi$ asymmetry) in the $\pi N$ Drell-Yan with TMD factorization at the kinematical configuration of the COMPASS experiments. We summarize the paper in Section VI.

\section{The TMD evolution of the distribution functions}
\label{Sec.TMDs_theo}

In this section, we present a review on the TMD evolution of the distribution functions. Particularly, we provide the evolution formalism for the unpolarized distribution function $f_1$, transversity $h_1$, Sivers function $f^\perp_{1}$ and the Boer-Mulders function $h_1^{\perp}$ of the proton, as well as $f_1$ and $h_1^\perp$ of the pion meson, within the Collins-11 TMD factorization scheme~\cite{Collins:2011zzd}.

In general, it is more convenient to solve the evolution equations for the TMD distributions in the coordinate space~($\bm{b}$ space) other than that in the transverse momentum $\bm{k}_\perp$ space, with $\bm{b}$ conjugate to $\bm{k}_\perp$ via Fourier transformation~\cite{Collins:1984kg,Collins:2011zzd}.
The TMD distributions $\tilde{F}(x,b;\mu,\zeta_F)$ in $\bm{b}$ space have two kinds of energy dependence, namely, $\mu$ is the renormalization scale related to the corresponding collinear PDFs, and $\zeta_F$ is the energy scale serving as a cutoff to regularize the light-cone singularity in the operator definition of the TMD distributions.
Here, $F$ is a shorthand for any TMD distribution function and the tilde denotes that the distribution is the one in $\bm{b}$ space.
If we perform the inverse Fourier Transformation on $\tilde{F}(x,b;\mu,\zeta_F)$, we recover the distribution function in the transverse momentum space $F_{q/H}(x,k_\perp;\mu,\zeta_F)$, which contains the information about the probability of finding a quark with specific polarization, collinear momentum fraction $x$ and transverse momentum $k_\perp$ in a specifically polarized hadron $H$.

\subsection{TMD evolution equations}

The energy evolution for the $\zeta_F$ dependence of the TMD distributions is encoded in the Collins-Soper~(CS)~\cite{Collins:1984kg,Idilbi:2004vb,Collins:2011zzd} equation:
\begin{align}
\frac{\partial\ \mathrm{ln} \tilde{F}(x,b;\mu,\zeta_F)}{\partial\ \sqrt{\zeta_F}}=\tilde{K}(b;\mu),
\end{align}
while the $\mu$ dependence is driven by the renormalization group equation as
\begin{align}
&\frac{d\ \tilde{K}}{d\ \mathrm{ln}\mu}=-\gamma_K(\alpha_s(\mu)),\\
&\frac{d\ \mathrm{ln} \tilde{F}(x,b;\mu,\zeta_F)}
{d\ \mathrm{ln}\mu}=\gamma_F(\alpha_s(\mu);{\frac{\zeta^2_F}{\mu^2}}),
\end{align}
with $\alpha_s$ the strong coupling at the energy scale $\mu$, $\tilde{K}$ the CS evolution kernel, and $\gamma_K$, $\gamma_F$ the anomalous dimensions.
The solutions of these evolution equations were studied in details in Refs.~\cite{Idilbi:2004vb,Collins:2011zzd,Collins:2014jpa}. Here, we will only discuss the final result.
The overall structure of the solution for $\tilde{F}(x,b;\mu,\zeta_F)$ is similar to that for the
Sudakov form factor. More specifically, the energy evolution of TMD distributions from an initial energy $\mu$ to another energy $Q$ is encoded in the Sudakov-like form factor $S$ by the exponential form $\mathrm{exp}(-S)$
\begin{equation}
\tilde{F}(x,b,Q)=\mathcal{F}\times e^{-S}\times \tilde{F}(x,b,\mu),
\label{eq:f}
\end{equation}
where $\mathcal{F}$ is the factor related to the hard scattering. Hereafter, we will set $\mu=\sqrt{\zeta_F}=Q$ and express $\tilde{F}(x,b;\mu=Q,\zeta_F=Q^2)$ as $\tilde{F}(x,b;Q)$.

As the $b$-dependence of the TMDs can provide very useful information regarding the transverse momentum dependence of the hadronic 3D structure through Fourier transformation, it is of fundamental importance to study the TMDs in $b$ space. In the small $b$ region, the $b$ dependence is perturbatively calculable, while in the large $b$ region, the dependence turns to be nonperturbative and may be obtained from the experimental data.
To combine the perturbative information at small $b$ with the nonperturbative part at large $b$, a matching procedure must be introduced with a parameter $b_{\mathrm{max}}$ serving as the boundary between the two regions. The prescription also allows
for a smooth transition from perturbative to nonperturbative
regions and avoids the Landau pole singularity in $\alpha_s(\mu_b)$.
A $b$-dependent function $b_\ast$ is defined to have the property $b_\ast\approx b$ at low values of $b$ and $b_{\ast}\approx b_{\mathrm{max}}$ at large $b$ values.
In this paper, we adopt the original CSS prescription~\cite{Collins:1984kg}:
 \begin{align}
 b_\ast=b/\sqrt{1+b^2/b_{\rm max}^2}  \ ,~b_{\rm max}<1/\Lambda_{\textrm{QCD}}.
\end{align}
The typical value of $b_{\mathrm{max}}$ is chosen around $1\ \mathrm{GeV}^{-1}$ to guarantee that $b_{\ast}$ is always in the perturbative region.
Besides the CSS prescription, there were several different prescriptions in literature.
In Refs.~\cite{Collins:2016hqq,Bacchetta:2017gcc} a function $b_{\rm min}(b)$ decreasing with increasing $1/Q$ was also introduced to match the TMD factorization with the fixed-order collinear calculations in the very small $b$ region.

In the small $b$ region $1/Q \ll b \ll 1/ \Lambda_{\textrm{QCD}}$, the TMD distributions at fixed energy $\mu$ can be expressed as the convolution of the perturbatively calculable coefficients and the corresponding collinear PDFs or the multiparton correlation functions~\cite{Collins:1981uk,Bacchetta:2013pqa}
\begin{equation}
\tilde{F}_{q/H}(x,b;\mu)=\sum_i C_{q\leftarrow i}\otimes F_{i/H}(x,\mu).
\label{eq:small_b_F}
\end{equation}
Here, $\otimes$ stands for the convolution in the momentum fraction $x$
\begin{equation}
 C_{q\leftarrow i}\otimes f_1^{i/H}(x,\mu)\equiv \int_{x}^1\frac{d\xi}{\xi} C_{q\leftarrow i}(x/\xi,b;\mu)f_1^{i/H}(\xi,\mu)
 \label{eq:otimes}
\end{equation}
and $f^{i/H}(x,\mu)$ is the corresponding collinear counterpart of flavor $i$ in hadron $H$ at the energy scale $\mu$.
The latter one could be a dynamic scale related to $b_\ast$ by $\mu_b=c_0/b_\ast$, with $c_0=2e^{-\gamma_E}$ and the Euler Constant $\gamma_E\approx0.577$~\cite{Collins:1981uk}.
The perturbative hard coefficients $C_{q\leftarrow i}$, independent of the initial hadron type, have been calculated for the parton-target case~\cite{Collins:1981uw,Aybat:2011zv} as the series of $(\alpha_s/\pi)$ and the results have been presented in Ref.~\cite{Bacchetta:2013pqa}~(see also Appendix A of Ref.~\cite{Aybat:2011zv}).

\subsection{Sudakov form factors for the proton and the pion}

The Sudakov-like form factor $S$ in Eq.~(\ref{eq:f}) can be separated into the perturbatively calculable part $S_{\mathrm{P}}$ and the nonperturbative part $S_{\mathrm{NP}}$
\begin{equation}
\label{eq:S}
S=S_{\mathrm{P}}+S_{\mathrm{NP}}.
\end{equation}
According to the studies in Refs.~\cite{Echevarria:2014xaa,Kang:2011mr,Aybat:2011ge,Echevarria:2012pw,Echevarria:2014rua},
the perturbative part of the Sudakov form factor $S_{P}$ has the same result among different kinds of distribution functions, i.e., $S_{P}$ is spin-independent.
It has the general form
\begin{equation}
\label{eq:Spert}
S_{\mathrm{P}}(Q,b_\ast)=\int^{Q^2}_{\mu_b^2}\frac{d\bar{\mu}^2}{\bar{\mu}^2}\left[A(\alpha_s(\bar{\mu}))
\mathrm{ln}\frac{Q^2}{\bar{\mu}^2}+B(\alpha_s(\bar{\mu}))\right].
\end{equation}
The coefficients $A$ and $B$ in Eq.(\ref{eq:Spert}) can be expanded as the series of $\alpha_s/{\pi}$:
\begin{align}
A=\sum_{n=1}^{\infty}A^{(n)}(\frac{\alpha_s}{\pi})^n,\\
B=\sum_{n=1}^{\infty}B^{(n)}(\frac{\alpha_s}{\pi})^n.
\end{align}
Here, we list $A^{(n)}$ to $A^{(2)}$ and $B^{(n)}$ to $B^{(1)}$ up to the accuracy of next-to-leading-logarithmic (NLL) order~\cite{Collins:1984kg,Landry:2002ix,Qiu:2000ga,Kang:2011mr,Aybat:2011zv,Echevarria:2012pw}:
\begin{align}
A^{(1)}&=C_F\\
A^{(2)}&=\frac{C_F}{2}\left[C_A\left(\frac{67}{18}-\frac{\pi^2}{6}\right)-\frac{10}{9}T_Rn_f\right]\\
B^{(1)}&=-\frac{3}{2}C_F.
\end{align}
For the nonperturbative form factor $S_{\mathrm{NP}}$, it can not be analytically calculated by the perturbative method, which means it has to be parameterized to obtain the evolution information in the nonperturbative region.

The general form of $S_{\rm NP}(Q;b)$ was suggested as~\cite{Collins:1984kg}
\begin{equation}
\label{eq:snp_gene}
S_{\rm NP}(Q;b)=g_2(b)\ln Q/Q_0 +g_1(b) \ .
\end{equation}
The nonperturbative functions $g_1(b)$ and $g_2(b)$ are functions of the impact parameter $b$ and depend on
the choice of $b_{\rm max}$. To be more specific, $g_2(b)$ contains the information on the large $b$ behavior of the evolution kernel $\tilde K$. Also, according to the power counting analysis in Ref.~\cite{Korchemsky:1994is}, $g_2(b)$ shall follow the power behavior as $b^2$ at small-$b$ region, which can be an essential constraint for the parametrization of $g_2(b)$.
The well-known Brock-Landry-Nadolsky-Yuan (BLNY) fit parameterizes $g_2(b)$ as $g_2 b^2$ with $g_2$ a free parameter~\cite{Landry:2002ix}.
We note that $g_2(b)$ is universal for
different types of TMDs and does not depend on the particular process, which is an important prediction of QCD factorization
theorems involving TMDs~\cite{Collins:2011zzd,Aybat:2011zv,Echevarria:2014xaa,Kang:2015msa}.
The nonperturbative function $g_1(b)$ contains information on the intrinsic nonperturbative transverse motion of bound partons, namely, it should depend on the type of hadron and the quark flavor as well as $x$ for TMD distributions. As for the TMD fragmentation functions, it may depend on $z_h$, the type of the produced hadron, and the quark flavor.
In other words, $g_1(b)$ depends on the specific TMDs.

There are several extractions for $S_{\rm NP}$ in literature, we review some often-used forms below.

The original BLNY fit parameterized $S_{\rm NP}$ as~\cite{Landry:2002ix}
\begin{equation}
\label{snp:BLNY}
\left(g_1+g_2\ln (Q/2Q_0)+g_1g_3\ln(100\,x_1x_2)\right)b^2,
\end{equation}
where $x_1$ and $x_2$ are the longitudinal momentum fractions of the incoming hadrons carried by
the initial state quark and antiquark.
The BLNY parameterization proved to be very reliable to describe Drell-Yan data and $W^\pm,Z$ boson production~\cite{Landry:2002ix}.
However, when the parametrization is extrapolated to the typical SIDIS kinematics in HERMES and COMPASS, the transverse momentum distribution of hadron can not be described by the BLNY-type fit~\cite{Sun:2013dya,Su:2014wpa}.

Inspired by Refs.~\cite{Landry:2002ix,Konychev:2005iy}, a widely used parametrization of $S_{\rm NP}$ for TMD distributions or fragmentation functions was proposed~\cite{Landry:2002ix,Konychev:2005iy,Davies:1984sp,Ellis:1997sc,Bacchetta:2013pqa,Echevarria:2014xaa}
\begin{align}
\label{snp:gaussian}
S_{\rm NP}^{\rm pdf/ff} &= b^2\left(g_1^{\rm pdf/ff}+ \frac{g_2}{2} \ln\frac{Q}{Q_0}\right),
\end{align}
where the factor $1/2$ in front of $g_2$ comes from the fact that only one hadron is involved for the parametrization of $S_{\rm NP}^{\rm pdf/ff}$, while the parameter in Ref.~\cite{Konychev:2005iy} is for $pp$ collisions.
The parameter $g_1^{\rm pdf/ff}$ in Eq.~(\ref{snp:gaussian}) depends on the type of TMDs, which can be regarded as the width of intrinsic
transverse momentum for the relevant TMDs at the initial energy scale $Q_0$~\cite{Qiu:2000ga,Aybat:2011zv,Anselmino:2012aa}.
Assuming a Gaussian form, one can obtain
\begin{align}
g_1^{\rm pdf} = \frac{\langle k_\perp^2\rangle_{Q_0}}{4},
\qquad
g_1^{\rm ff} = \frac{\langle p_T^2\rangle_{Q_0}}{4z^2},
\end{align}
where $\langle k_\perp^2\rangle_{Q_0}$ and $\langle p_T^2\rangle_{Q_0}$
represent the relevant averaged intrinsic transverse momenta squared for TMD distributions and TMD fragmentation functions at the initial scale $Q_0$, respectively.

Since the original BLNY fit fails to
simultaneously describe Drell-Yan process and SIDIS process, in Ref.~\cite{Su:2014wpa} the authors proposed a new form for $S_{\rm NP}$ which releases the tension between the BLNY fit to the Drell-Yan~(such as $W$, $Z$ and low energy
Drell-Yan pair productions) data
and the fit to the SIDIS data from HERMES/COMPASS in the CSS resummation formalism.
In addition, the $x$-dependence in Eq.~(\ref{snp:BLNY}) was separated with a power law behavior assumption: $(x_0/x)^\lambda$, where $x_0$ and $\lambda$ are the fixed parameters as $x_0=0.01$ and $\lambda=0.2$.
The two different behaviors (logarithmic in Eq.~(\ref{snp:BLNY}) and power law) will differ
in the intermediate $x$ regime. Ref.~\cite{Sun:2013dya} showed that a direct integration of the evolution kernel from low $Q$ to high $Q$ led to the form of $\ln(Q)$ term as $\ln(b/b_\ast)\ln(Q)$ and could describe the SIDIS and Drell-Yan data with $Q$ values ranging from a few GeV
to 10 GeV.
Thus, the $g_2(b)$ term was modified to the form of $\ln(b/b_\ast)$ and the functional form of $S_{\rm NP}$ extracted in Ref.~\cite{Su:2014wpa} turned to the form
\begin{equation}
g_1b^2+g_2\ln(b/b_*)\ln(Q/Q_0)+g_3b^2\left((x_0/x_1)^\lambda+(x_0/x_2)^\lambda\right)\ .
\end{equation}
At small $b$ region~($b$ is much smaller than $b_{\rm max}$), the parametrization of the $g_2(b)$ term $g_2\ln(b/b_*)$ can be approximated as $b^2/(2b^2_{\rm max})$, which satisfied the constraint of the $b^2$ behavior for $g_2(b)$. However, at large $b$ region, the logarithmic behavior will lead to different predictions on the $Q^2$ dependence, since the Gaussian-type parametrization suggests that it is strongly suppressed~\cite{Aidala:2014hva}.
This form has been suggested in an early research by Collins and Soper~\cite{Collins:1985xx}, but has not yet
been adopted in any phenomenological study until the study in Ref.~\cite{Su:2014wpa}.
The comparison between the original BLNY parametrization and this form with the experimental data of Drell-Yan type process has shown that the new form of $S_{\rm NP}$ can fit with the data as equally well as the original BLNY parametrization.

In Ref.~\cite{Bacchetta:2017gcc}, the $g_2(b)$ function was parameterized as $g_2 b^2$, following the BLNY convention. Furthermore, in the function $g_1(b)$, the Gaussian width also depends on $x$. The authors simultaneously fit the experimental data of SIDIS process from HERMES and COMPASS Collaborations, the Drell-Yan events at low energy, and the $Z$ boson production with totally 8059 data points. The extraction can describe the data well in the regions where TMD factorization is supposed to hold.

\begin{figure}
\centering
\includegraphics[width=0.8\columnwidth]{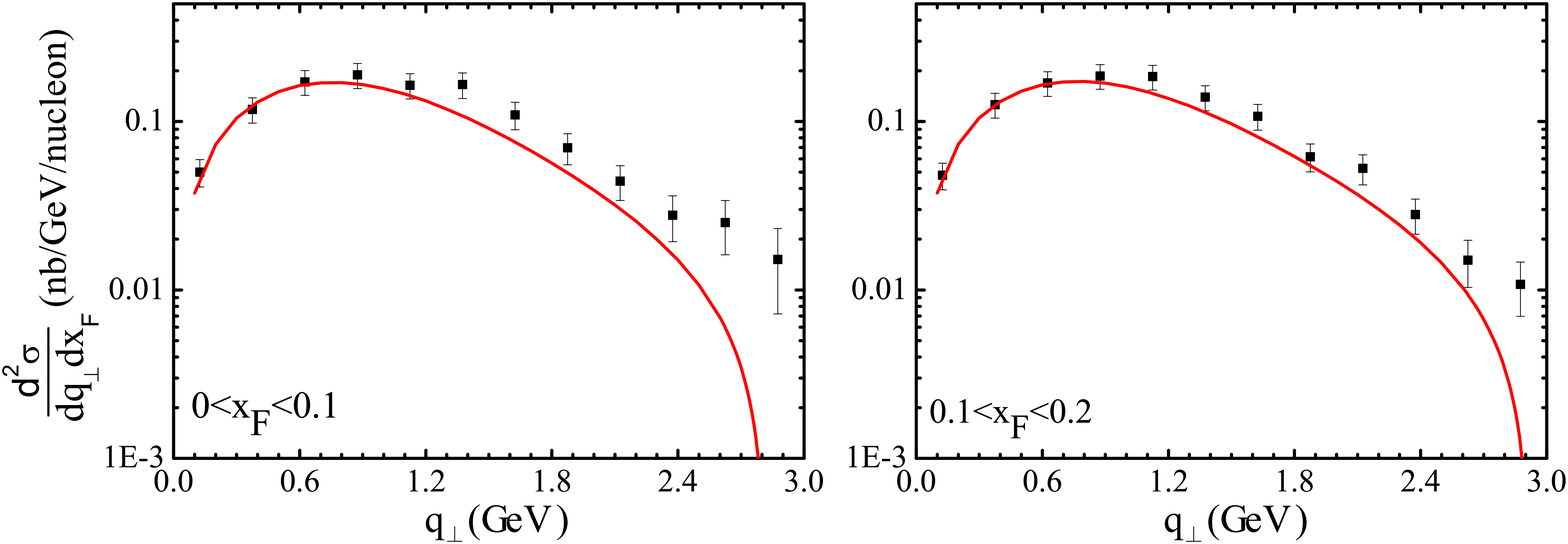}\\
\includegraphics[width=0.8\columnwidth]{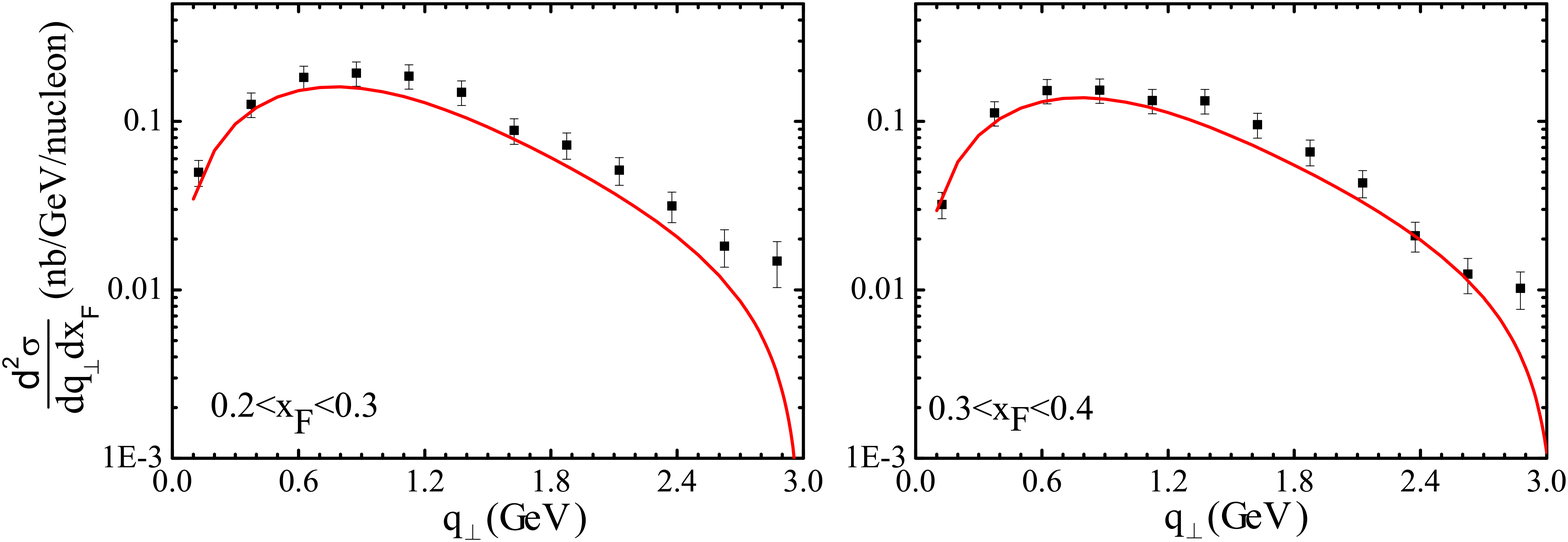}\\
\includegraphics[width=0.8\columnwidth]{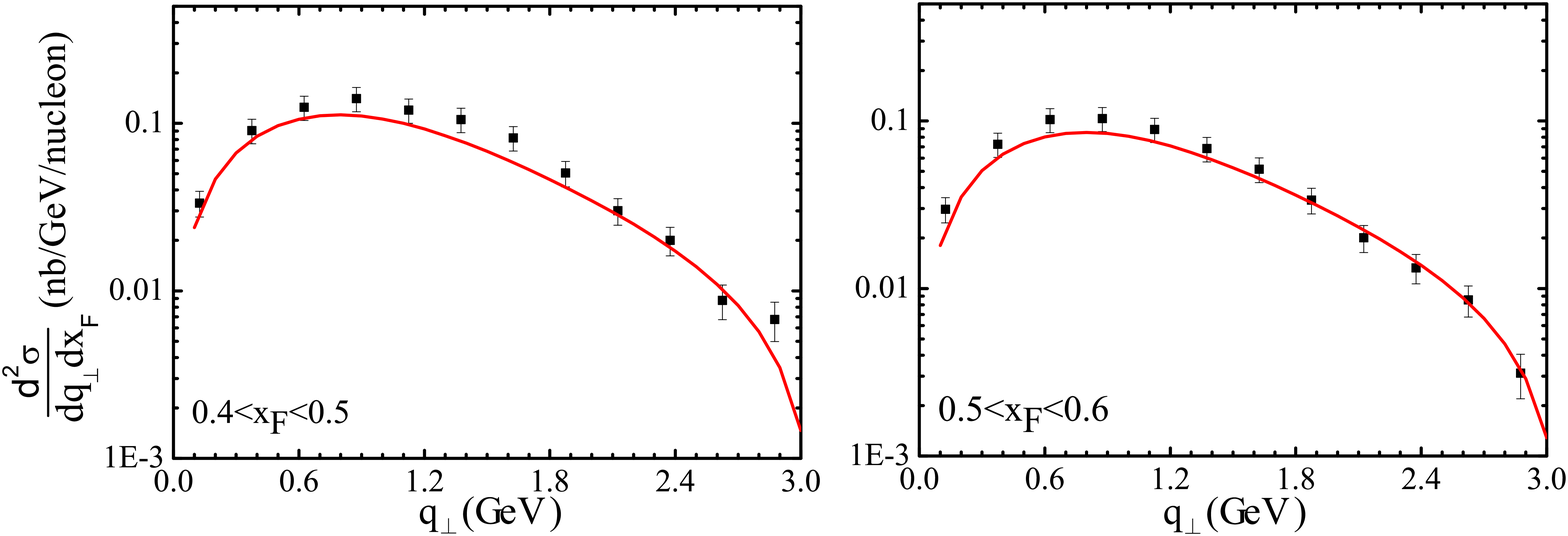}\\
\includegraphics[width=0.8\columnwidth]{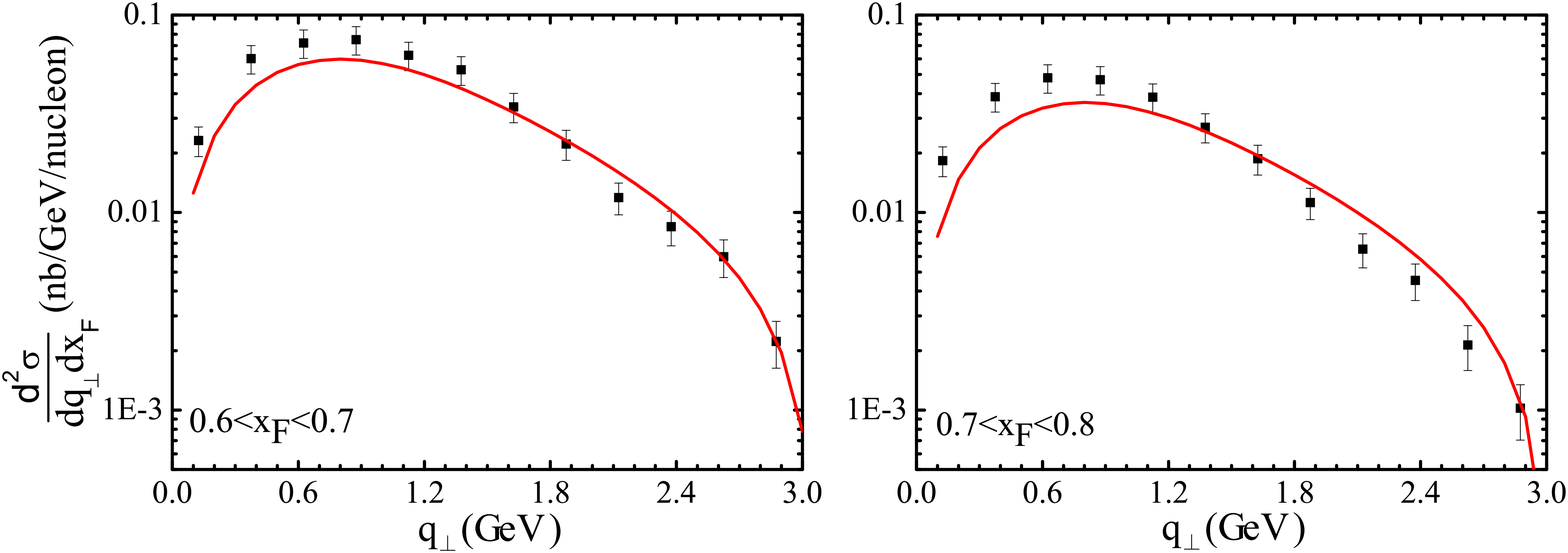}
\caption{The fitted cross section (solid line) of pion-nucleon Drell-Yan as functions of $q_\perp$, compared with the E615 data (full squre), for different $x_F$ bins in the range $0<x_F<0.8$. The error bars shown here include the statistical error and the $16\%$ systematic error. Figure from Ref.~\cite{Wang:2017zym}.}
\label{fig:comparison1}
\end{figure}

To study the pion-nucleon Drell-Yan data, it is also necessary to know the nonperturbative Sudakov form factor for the pion meson.
In Ref.~\cite{Wang:2017zym}, we extended the functional form for the proton TMDs~\cite{Su:2014wpa} to the case of the pion TMDs:
\begin{align}
S^{f_1^{q/\pi}}_{\mathrm{NP}}=g^{q/\pi}_1b^2+g^{q/\pi}_2\mathrm{ln}\frac{b}{b_{\ast}}\mathrm{ln}\frac{Q}{Q_0},
\label{eq:snppi}
\end{align}
with $g^{q/\pi}_1$ and $g^{q/\pi}_2$ the free parameters.
Adopting the functional form of $S_{\textrm NP}$ in Eq.~(\ref{eq:snppi}), for the first time, we performed the extraction~\cite{Wang:2017zym} of the nonperturbative Sudakov form factor for the unpolarized TMD PDF of pion meson using the experimental data in the $\pi^-\,p$ Drell-Yan process collected by the E615 Collaboration at Fermilab~\cite{Conway:1989fs,Stirling:1993gc}.
The data fitting was performed by the package {\sc{MINUIT}}~\cite{James:1975dr,James:1994vla}, through a least-squares fit:
\begin{align}
\chi^2(\bm{\alpha})=\sum_{i=1}^{M}\sum_{j=1}^{N_i} \frac{(\mathrm{theo}(q_{\perp ij},\bm{\alpha})-\mathrm{data}_{ij})^2}{\mathrm{err}_{ij}^2}.
\label{eq:chisquare}
\end{align}
The total number of data in our fit is $N=\sum_i^{8}N_i=96$.
Since the TMD formalism is valid in the region $q_\perp\ll Q $, we did a simple data selection by removing the data in the region $q_\perp>3\ \mathrm{GeV}$.
We performed the fit by minimizing the chisquare in Eq.~(\ref{eq:chisquare}), and we obtained the following values for the two parameters:
\begin{align}
g^{q/\pi}_1=0.082 \pm 0.022 ,\quad g^{q/\pi}_2=0.394 \pm 0.103, \label{eq:gpi12}
\end{align}
with $\chi^2/\textrm{d.o.f} = 1.64$.

Fig.~\ref{fig:comparison1} plots the $q_\perp$-dependent differential cross section (solid line) calculated from the fitted values for $g^{q/\pi}_1$ and $g^{q/\pi}_2$ in Eq.~(\ref{eq:gpi12}) at the kinematics of E615 at different $x_F$ bins. The full squares with error bars denote the E615 data for comparison.
As Fig.~\ref{fig:comparison1} demonstrates, a good fit is obtained in the region $x_F<0.8$.

From the fitted result, we find that the value of the parameter $g_1^{q/\pi}$ is smaller than the parameter $g_1^{q/p}$ extracted in Ref.~\cite{Su:2014wpa} which used the same parameterized form. For the parameter $g_2^{q/\pi}$ we find that its value is very close to that of the parameter $g_2^{q/p}$ for the proton~\cite{Su:2014wpa} (here $g_2^{q/p} = g_2/2 =0.42$).
This may confirm that $g_2$ should be universal, e.g., $g_2$ is independent on the hadron type. Similar to the case of the proton, for the pion meson $g_2^\pi$ is several times larger than $g_1^\pi$.
We note that a form of $S^{f_{1,q/\pi}}_{\mathrm{NP}}$ motivated by the NJL model was given in Ref.~\cite{Ceccopieri:2018nop}.

\subsection{Solutions for different TMDs}

After solving the evolution equations and incorporating the Sudakov form factor, the scale-dependent TMD distribution function $\tilde{F}$ of the proton and the pion in $b$ space can be rewritten as
\begin{align}
\label{eq:tildeF}
\tilde{F}_{q/p}(x,b;Q)=e^{-\frac{1}{2}S_{\mathrm{P}}(Q,b_\ast)-S^{F_{q/p}}_{\mathrm{NP}}(Q,b)}
\mathcal{F}(\alpha_s(Q))\sum_i C_{q\leftarrow i}\otimes F_{i/p}(x,\mu_b),\\
\tilde{F}_{q/\pi}(x,b;Q)=e^{-\frac{1}{2}S_{\mathrm{P}}(Q,b_\ast)-S^{F_{q/\pi}}_{\mathrm{NP}}(Q,b)}
\mathcal{F}(\alpha_s(Q))\sum_i C_{q\leftarrow i}\otimes F_{i/\pi}(x,\mu_b).
\end{align}
Here, $F_{i/H}(x,\mu_b)$ is the corresponding collinear distributions at the initial energy scale $\mu_b$.
To be more specific, for the unpolarized distribution function $f_{1,q/H}$ and transversity distribution function $h_{1,q/H}$, the collinear distributions $F_{i/H}(x,\mu_b)$ are the integrated distribution functions $f_{1,q/H}(x,\mu_b)$ and $h_{1,q/H}(x,\mu_b)$. As for the Boer-Mulders function and Sivers function, the collinear distributions are the corresponding multi-parton correlation functions.
Thus, the unpolarized distribution function of the proton and pion in $b$ space can be written as
\begin{align}
\label{eq:unpo_f}
&\tilde{f}_{1,q/p}(x,b;Q)=e^{-\frac{1}{2}S_{\mathrm{pert}}(Q,b_\ast)-S^{f_{1,q/p}}_{\mathrm{NP}}(Q,b)}\mathcal{F}(\alpha_s(Q))
\sum_iC_{q\leftarrow i}\otimes f_{1,i/p}(x,\mu_b)\\
&\tilde{f}_{1,q/\pi}(x,b;Q) =e^{-\frac{1}{2}S_{\mathrm{pert}}(Q,b_\ast)-S^{f_{1,q/\pi}}_{\mathrm{NP}}(Q,b)}
\mathcal{F}(\alpha_s(Q))\sum_iC_{q\leftarrow i}\otimes f_{1,i/\pi}(x,\mu_b).\label{eq:unpo_fpi}
\end{align}
If we perform a Fourier transformation on the $\tilde{f}_{1,q/H}(x,b;Q)$, we can obtain the distribution function in $k_\perp$ space as
\begin{align}
&f_{1,q/p}(x,k_\perp;Q)=\int_0^\infty\frac{dbb}{2\pi}J_0(k_\perp b)\tilde{f}_{1,q/p}(x,b;Q),\\
&f_{1,q/\pi}(x,k_\perp;Q)=\int_0^\infty\frac{dbb}{2\pi}J_0(k_\perp b)\tilde{f}_{1,q/\pi}(x,b;Q).
\end{align}
where $J_0$ is the Bessel function of the first kind, and $k_\perp = |\bm k_\perp|$.

Similarly, the evolution formalism of the proton transversity distribution in $b$ space and $k_\perp$-space can be obtained as~\cite{Kang:2015msa}
\begin{align}
&\tilde h_{1,q/p}(x,b;Q)=e^{-\frac{1}{2}{S}_{\rm P}(Q,b_\ast)-S_{\rm NP}^{h_1}(Q,b)} \, \mathcal{H}(\alpha_s(Q))\,  \sum_i\delta C_{q\gets i}\otimes h_{1,i/p}(x,\mu_b),\\
&h_{1,q/p}(x,k_\perp;Q)=\int_0^\infty\frac{dbb}{2\pi}J_0(k_\perp b)\tilde{h}_{1,q/p}(x,b;Q), \label{eq:trans}
\end{align}
where $\mathcal{H}$ is the hard factor, and $\delta C_{q\gets i}$ is the coefficient convoluted with the transversity. The TMD evolution formalism in Eq.~(\ref{eq:trans}) has been applied in Ref.~\cite{Kang:2015msa} to extract the transversity distribution from the SIDIS data.

The Sivers function and Boer-Mulders function, which are T-odd, can be expressed as follows in $b$-space~\cite{Echevarria:2014xaa}
\begin{align}
&\tilde{f}_{1T,q/H}^{\perp\alpha(\mathrm{DY})}(x,b;\mu,\zeta_F)=\int d^2\bm{k}_\perp e^{-i\vec{\bm{k}}_\perp\cdot\vec{\bm{b}}}\frac{k^\alpha_\perp}{M_p}
f^{\perp(\mathrm{DY})}_{1T,q/H}(x,\bm{k}_\perp;\mu),\nonumber\\
&\tilde{h}_{1,q/H}^{\perp\alpha(\mathrm{DY})}(x,b;\mu,\zeta_F)=\int d^2\bm{k}_\perp e^{-i\vec{\bm{k}}_\perp\cdot\vec{\bm{b}}}\frac{k^\alpha_\perp}{M_p}
h^{\perp(\mathrm{DY})}_{1,q/H}(x,\bm{k}_\perp;\mu).
\end{align}
Here, the superscript ``DY" represents the distributions in the Drell-Yan process. Since QCD predicts that the sign of the distributions changes in the SIDIS process and Drell-Yan process, for the distributions in SIDIS process, there has to be an extra minus sign regard to $f^{\perp(\mathrm{DY})}_{1T,q/H}$ and $h^{\perp(\mathrm{DY})}_{1,q/H}$.

Similar to what has been done to the unpolarized distribution function and transversity distribution function,
in the low $b$ region, the Sivers function $\tilde{f}_{1T,q/H}^{\perp\alpha(\mathrm{DY})}$ can also be expressed as the convolution of perturbatively calculable hard coefficients and the corresponding collinear correlation functions as~\cite{Sun:2013hua,Kang:2011mr}
\begin{equation}
\tilde{f}_{1T,q/H}^{\perp\alpha(\mathrm{DY})}(x,b;\mu)=(\frac{-ib^\alpha}{2})\sum_i \Delta C^T_{q\leftarrow i}\otimes f^{(3)}_{i/p}(x',x'';\mu).
\label{eq:Siv_fixed_engy}
\end{equation}
Here, $f^{(3)}_{i/p}(x',x'')$ denotes the twist-three quark-gluon-quark or trigluon correlation
functions, among which the transverse spin-dependent Qiu-Sterman matrix element $T_{q,F}(x',x'')$~\cite{Qiu:1991pp,Qiu:1991wg,Qiu:1998ia} is the most relevant one.
Assuming that the Qiu-Sterman function $T_{q,F}(x,x)$ is the main contribution, the Sivers function in $b$-space becomes
\begin{align}
\label{eq:Siv_b}
&\tilde{f}_{1T,q/H}^{\perp \alpha(\mathrm{DY})}(x,b;Q)=(\frac{-ib^\alpha}{2})\mathcal{F}_{\rm Siv}(\alpha_s(Q))\sum_i \Delta C^T_{q\leftarrow
i}\otimes T_{i/H,F}(x,x;\mu_b) e^{-S^{\mathrm{siv}}_{\mathrm{NP}}-
\frac{1}{2}S_{\mathrm{P}}},
\end{align}
where $\mathcal{F}_{\rm Siv}$ is the factor related to the hard scattering.
The Boer-Mulders function in $b$-space follows the similar result for the Sivers function as:
\begin{align}
\label{eq:BM_b}
\widetilde{h}_{1,q/H}^{\alpha\perp(\mathrm{DY})}(x,b;Q)=(\frac{-ib^\alpha}{2})
\mathcal{H}_{\rm BM}(\alpha_s(Q))
\sum_i C^{\rm BM}_{q\leftarrow
i}\otimes T^{(\sigma)}_{i/H,F}(x,x;\mu_b)e^{-S^{\rm BM}_{\mathrm{NP}}-\frac{1}{2}S_{\mathrm{P}}},
\end{align}
Here, $C^{\rm BM}_{q\leftarrow
i}$ stands for the flavor-dependent hard coefficients convoluted with $T^{(\sigma)}_{i/H,F}$, $\mathcal{H}_{\rm BM}$ the hard scattering factor and
$T^{(\sigma)}_{i/H,F}(x,x;\mu_b)$ denotes the chiral-odd twist-3 collinear correlation function.
After performing the Fourier transformation back to the transverse momentum space, one can get the Sivers function and the Boer-Mulders function as
\begin{align}
&\frac{k_\perp}{M_H}f_{1T,q/H}^{\perp}(x,k_\perp;Q)=\int_0^\infty db
(\frac{b^2}{2\pi})J_1(k_\perp b)\mathcal{F}_{\rm Siv}(\alpha_s(Q))\sum_i \Delta C^T_{q\leftarrow i}\otimes
f^{\perp(1)}_{1T,i/H}(x,\mu_b) e^{-S^{\mathrm{siv}}_{\mathrm{NP}}-
\frac{1}{2}S_{\mathrm{P}}},\label{eq:sivers}\\
&\frac{k_\perp}{M_H}h^\perp_{1,q/H}(x,k_\perp;Q)=\int_0^\infty db(\frac{b^2}{2\pi})J_1(k_\perp b)\mathcal{H}_{\rm BM}(\alpha_s(Q))\sum_i  C^{\rm BM}_{q\leftarrow i}\otimes
h^{\perp(1)}_{1,i/H}(x;\mu_b)e^{-S^{\rm BM}_{\mathrm{NP}}-\frac{1}{2}S_{\mathrm{P}}},
\label{eq:BM_kt}
\end{align}
and $T_{q,F}(x,x;\mu_b)$ and $T^{(\sigma)}_{i/H,F}(x,x;\mu_b)$ are related to Sivers function and Boer-Mulders function as~\cite{Sun:2013hua,Kang:2011mr}
\begin{align}
&T_{q/H,F}(x,x)=\int d^2k_\perp \frac{|k_\perp^2|}{M_H}f^{\perp{\textrm{(DY)}}}_{1T,q/H}(x,k_\perp) = 2M_H\,f_{1T,q/H}^{\perp(1) \textrm{(DY)}}(x), \\
&T^{(\sigma)}_{q/H,F}(x,x)=\int d^2k_\perp\frac{|k_\perp^2|}{M_H}h_{1,q/H}^{\perp \textrm{(DY)}}(x,k_\perp)
= 2M_H h_{1,q/H}^{\perp (1) \textrm{(DY)}}(x).
\end{align}

The TMD evolution formalism in Eq.~(\ref{eq:sivers}) has been applied to extract~\cite{Aybat:2011ge,Anselmino:2012aa,Anselmino:2012re,Echevarria:2014xaa,Boglione:2018dqd} the Sivers function.
The similar formalism in Eq.~(\ref{eq:BM_kt}) could be used to improvel the previous extractions of the proton Boer-Mulders function~\cite{Zhang:2008nu,Lu:2009ip,Barone:2009hw,Barone:2010gk} and future extraction of the pion Boer-Mulders function.

\section{Physical observables in $\pi N$ Drell-Yan process within TMD factorization}

\label{Sec.DY_theo}
In this section we will set up the necessary framework for physical observables in $\pi$-$N$ Drell-Yan process within TMD factorization by considering the evolution effects of the TMD distributions, following the procedure developed in Ref.~\cite{Collins:2011zzd}.

In Drell-Yan process
\begin{equation}
\label{eq:DYprocess}
H_A(P_\pi)+H_B(P_N)\longrightarrow \gamma^*(q)+X \longrightarrow l^+(\ell)+l^-(\ell')+X,
\end{equation}
$P_{\pi/N}$ and $q$ denote the momenta of the incoming hadron $\pi/N$ and the virtual photon, respectively; $q$ is a time-like vector, namely, $Q^2=q^2>0$, which is the invariant mass square of the final-state lepton pair.
One can define the following useful kinematical variables to express the cross section:
\begin{align}
&s=(P_\pi+P_N)^2,\quad x_{\pi/N}=\frac{Q^2}{2P_{\pi/N}\cdot q},
x_F=2q_L/s=x_\pi-x_N,\nonumber\\ &\tau=Q^2/s=x_\pi x_N,\quad y=\frac{1}{2}\mathrm{ln}\frac{q^+}{q^-}=\frac{1}{2}\mathrm{ln}\frac{x_\pi}{x_N},
\end{align}
where $s$ is the center-of-mass energy squared; $x_{\pi/N}$ is the light-front momentum fraction carried by the annihilating quark/antiquark in the incoming hadron $\pi/N$; $q_L$ is the longitudinal momentum of the virtual photon in the c.m. frame of the incident hadrons; $x_F$ is the Feynman $x$ variable, which corresponds to the longitudinal momentum fraction carried by the lepton pair; and $y$ is the rapidity of the lepton pair. Thus, $x_{\pi/N}$ is expressed as the function of $x_F$, $\tau$ and of $y$, $\tau$
\begin{align}
x_{\pi/N}=\frac{\pm x_F+\sqrt{x_F^2+4\tau}}{2},\quad x_{\pi/N}=\sqrt{\tau} e^{\pm y}.
\end{align}

\subsection{Differential cross section for unpolarized Drell-Yan process}

The differential cross section formulated in TMD factorization is usually expressed in the $b$-space to
guarantee conservation of the transverse momenta of the emitted soft gluons. Later on it can be transformed back to the transverse momentum space to represent the experimental observables.
We will introduce the physical observables in the following part of this section.

The general differential cross section for the unpolarized Drell-Yan process can be written as~\cite{Collins:1984kg}
\begin{equation}
\label{eq:dsigma_UU}
\frac{d^4\sigma_{UU}}{dQ^2dyd^2\bm{q}_{\perp}}=\sigma_0\int \frac{d^2b}{(2\pi)^2}e^{i\vec{\bm{q}}_{\perp}\cdot \vec{\bm{b}}}\widetilde{W}_{UU}(Q;b)+Y_{UU}(Q,q_{\perp})
\end{equation}
where $\sigma_0=\frac{4\pi\alpha_{em}^2}{3N_CsQ^2}$ is the cross section at tree level with $\alpha_{em}$ the fine-structure constant, $\widetilde{W}(Q;b)$ is the structure function in the $b$-space which contains all-order resummation results and dominates in the low $q_{\perp}$ region ($q_{\perp}\ll Q$); and the $Y$ term provides necessary correction at $q_{\perp}\sim Q$. In this work we will neglect the $Y$-term, which means that we will only consider the first term on the r.h.s of Eq.~(\ref{eq:dsigma_UU}).

In general, TMD factorization~\cite{Collins:2011zzd} aims at separating well defined TMD distributions such that they can be used in different processes through a universal way and expressing the scheme/process dependence in the corresponding hard factors. Thus, $\widetilde{W}(Q;b)$ can be expressed as ~\cite{Prokudin:2015ysa}
\begin{align}
\label{eq:WUU_1}
\widetilde{W}_{UU}(Q;b)=H_{UU}(Q;\mu) \sum_{q,\bar{q}}e_q^2\tilde{f}_{q/\pi}^\mathrm{sub}(x_\pi,b;\mu,\zeta_F)
\tilde{f}_{q/p}^\mathrm{sub}(x_p,b;\mu,\zeta_F),
\end{align}
where $\tilde{f}_{q/H}^\mathrm{sub}$ is the subtracted distribution function in the $b$ space and $H_{UU}(Q;\mu)$ is the factor associated with hard scattering. The superscript ``sub" represents the distribution function with the soft factor subtracted.
The subtraction guarantees the absence of light-cone singularities in the TMDs and the self-energy divergencies of the
soft factors~\cite{Collins:1981uk,Collins:2011zzd}.
However, the way to subtract the soft factor in the distribution function and the hard factor $H_{UU}(Q;\mu)$ depends on the scheme to regulate the light-cone singularity in the TMD definition~\cite{Collins:1981uk,
Collins:1984kg,Ji:2004wu,Collins:2004nx,Collins:2011zzd,Mantry:2009qz,Becher:2010tm,
GarciaEchevarria:2011rb,Chiu:2012ir,Ji:2014hxa}, leading to the scheme dependence in the TMD factorization.
In literature, several different schemes are used~\cite{Prokudin:2015ysa}: the CSS scheme ~\cite{Collins:1981uk,Collins:1984kg}, the Collins-11 (JCC) scheme~\cite{Collins:2011zzd}, the Ji-Ma-Yuan (JMY) scheme~\cite{Ji:2004wu,Ji:2004xq}, and the lattice scheme~\cite{Ji:2014hxa}. Although different schemes are adopted, the final results of the structure functions $\widetilde{W}(Q;b)$ as well as the differential cross section should not depend on a specific scheme. In the following we will apply the JCC and JMY schemes to display the scheme-independence of the unpolarized differential cross section.

The hard $H_{UU}(Q;\mu)$ have different forms in the JCC and JMY schemes:
\begin{align}
&H^{\mathrm{JCC}}(Q;\mu)=1+\frac{\alpha_s(\mu)}{2\pi}C_F
\left(3\ln\frac{Q^2}{\mu^2}-\ln^2\frac{Q^2}{\mu^2}+\pi^2-8\right), \\
&H^{\mathrm{JMY}}(Q;\mu,\rho)=1+\frac{\alpha_s(\mu)}{2\pi}C_F
\left((1+\ln\rho^2)\ln\frac{Q^2}{\mu^2}-\ln\rho^2+\ln^2\rho+2\pi^2-4\right).
\end{align}
Like $\zeta_F$, here $\rho$ is another variable to regulate the light-cone singularity of TMD distributions. The scheme dependence of the distribution function is manifested in the hard factor $\mathcal{F}(\alpha_s(Q))$, which has the following forms in different schemes:
\begin{align}
\label{eq:f_factor}
&\tilde{\mathcal{F}}^{\mathrm{JCC}}(\alpha_s(Q))=1+\mathcal{O}(\alpha_s^2), \\ &\tilde{\mathcal{F}}^{\mathrm{JMY}}(\alpha_s(Q),\rho)=1+\frac{\alpha_s}{2\pi}C_F
\left[\mathrm{ln}\rho-\frac{1}{2}\mathrm{ln}^2\rho-\frac{\pi^2}{2}-2\right],
\end{align}
The $C$ coefficients in Eqs.~(\ref{eq:unpo_f}) and (\ref{eq:unpo_fpi}) do not depend on the types of initial hadrons and are calculated for the parton-target case~\cite{Collins:1981uw,Aybat:2011zv} with the results presented in Ref.~\cite{Bacchetta:2013pqa} (see also Appendix A of Ref.~\cite{Aybat:2011zv})
\begin{align}
\label{eq:cfactor}
&C_{q\leftarrow q^{\prime}}(x,b;\mu,\zeta_F)=\delta_{qq^{\prime}}\left[\delta(1-x)+\frac{\alpha_s}{\pi}\left(\frac{C_F}{2}(1-x)\right)\right],\\
&C_{q\leftarrow g}(x,b;\mu,\zeta_F)=\frac{\alpha_s}{\pi}T_Rx(1-x),
\end{align}
where $C_F=(N_C^2-1)/(2N_C)$, $T_R=1/2$.

One can absorb the scheme-dependent hard factors $H_{UU}(Q;\mu)$ and $\mathcal{F}$ of the TMD distributions into the $C$-functions using
\begin{align}
C_{j\leftarrow i}^\prime = C_{j\leftarrow i}\times \mathcal{F}\times\sqrt{H_{UU}(Q;\mu=Q)}.
\end{align}
The results for the splitting to quark are
\begin{align}
&C^\prime_{q\leftarrow q^{\prime}}(x,b;\mu_b)=\delta_{qq^{\prime}}\left[\delta(1-x)+\frac{\alpha_s}{\pi}
\left(\frac{C_F}{2}(1-x)+\frac{C_F}{4}(\pi^2-8)\delta(1-x)\right)\right],\\
&C^\prime_{q\leftarrow g}(x,b;\mu_b)=\frac{\alpha_s}{\pi}T_Rx(1-x).
\end{align}
The new $C$-coefficients turn out to be scheme independent (independent on $\rho$) ~\cite{Catani:2000vq} but process dependent ~\cite{Nadolsky:1999kb,Koike:2006fn}.

With the new $C$-coefficients in hand, one can obtain the structure functions $\widetilde{W}_{UU}(Q;b)$ in $b$-space as
\begin{align}
\label{eq:wuu}
\widetilde{W}_{UU}(Q;b)&=e^{-S_\mathrm{pert}(Q^2,b)-
S^{f^{q/\pi}_{1}}_\mathrm{NP}(Q^2,b)-S^{f^{q/p}_{1}}_\mathrm{NP}(Q^2,b)}\nonumber\\
&\times \sum_{q,\bar{q}}e_q^2\,C^\prime_{q\leftarrow i}\otimes f_{i/\pi^-}(x_1,\mu_b)
\,C^\prime_{\bar{q}\leftarrow j}\otimes f_{j/p}(x_2,\mu_b).
\end{align}
After performing the Fourier transformation, we can get the differential cross section as
\begin{align}
\label{eq:dsig_UU_final}
&\frac{d^4\sigma}{dQ^2dyd^2\bm{q}_{\perp}}=\sigma_0\int_0^\infty \frac{db b}{2\pi}J_0(q_\perp b)\times \widetilde{W}_{UU}(Q;b),
\end{align}
where $J_0$ is the zeroth order Bessel function of the first kind.

\subsection{The Sivers asymmetry}

In the Drell-Yan process with a
$\pi$ beam colliding on the transversely polarized nucleon target, an important physical observable is the Sivers asymmetry, as it can test the sign change of the Sivers function between SIDIS and Drell-Yan processes, a fundamental prediction in QCD.
The future precise measurement of the Sivers asymmetry in $\pi N$ Drell-Yan in a wide kinematical region can be also used to extract the Sivers function.
The Sivers asymmetry is usually defined as ~\cite{Echevarria:2014xaa}
\begin{align}
A_{UT}=\frac{d^4\Delta\sigma}{dQ^2dyd^2\bm{q}_{\perp}}\bigg{/}{\frac{d^4\sigma}{dQ^2dyd^2\bm{q}_{\perp}}},
\label{eq:asy_Sivers}
\end{align}
where $\frac{d^4\sigma}{dQ^2dyd^2\bm{q}_{\perp}}$ and $\frac{d^4\Delta\sigma}{dQ^2dyd^2\bm{q}_{\perp}}$ are the spin-averaged~(unpolarized) and spin-dependent differential cross section, respectively.
The latter one has the general form in the TMD factorization~\cite{Kang:2011mr,Echevarria:2014xaa,Sun:2013hua}
\begin{equation}
\label{eq:dsigma_UT}
\frac{d^4\Delta\sigma}{dQ^2dyd^2\bm{q}_{\perp}}=\sigma_0\epsilon_\perp^{\alpha\beta} S^\alpha_\perp\int \frac{d^2b}{(2\pi)^2}e^{i\vec{\bm{q}}_{\perp}\cdot \vec{\bm{b}}}\widetilde{W}^\beta_{UT}(Q;b)+Y^\beta_{UT}(Q,q_{\perp}).
\end{equation}
Similar to Eq.~(\ref{eq:dsigma_UU}), $\widetilde{W}_{UT}(Q,b)$ denotes the spin-dependent structure function in the $b$-space and dominates at $q_{\perp}\ll Q$, and $Y^\beta_{UT}$ provides correction for the single polarized process at $q_{\perp}\sim Q$.
The antisymmetric tensor $\epsilon^{\alpha\beta}_\perp$ is defined as $\epsilon^{\alpha\beta\mu\nu}P_\pi^\mu P_p^\nu / P_\pi\cdot P_p$, and $S_\perp$ is the transverse polarization vector of the proton target.

The structure function $\widetilde{W}_{UT}(Q,b)$ can be written in terms of the unpolarized distribution function of pion and Sivers function of proton as
\begin{align}
\label{eq:WUT}
&\widetilde{W}^{\alpha}_{UT}(Q;b)=H_{UT}(Q;\mu) \sum_{q,\bar{q}}e_q^2\tilde{f}_{1\, \bar{q}/\pi}(x_\pi,b;\mu,\zeta_F)
\tilde{f}_{1T\,q/p}^{\perp\alpha(\mathrm{DY})}(x_p,b;\mu,\zeta_F),
\end{align}
with $\tilde{f}_{1T\,q/p}^{\perp\alpha(\mathrm{DY})}(x_p,b;\mu,\zeta_F)$ given in Eq.~(\ref{eq:Siv_b}).
Similar to the unpolarized case, the scheme-dependent hard factors can be absorbed into the $C$-coefficients, leading to~\cite{Kang:2011mr,Sun:2013hua}
\begin{align}
\Delta C^T_{q\leftarrow q^{\prime}}(x,b;\mu_b)=\delta_{qq^{\prime}}\left[\delta(1-x)+\frac{\alpha_s}{\pi}
\left(-\frac{1}{4N_c}(1-x)+\frac{C_F}{4}(\pi^2-8)\delta(1-x)\right)\right].
\label{eq:cfactor_Siv}
\end{align}

The spin-dependent differential cross section in Eq.~(\ref{eq:dsigma_UT}) thus has the form
\begin{align}
\label{eq:dsig_UT_final}
\frac{d^4\Delta\sigma}{dQ^2dyd^2\bm{q}_{\perp}}=\frac{\sigma_0}{4\pi}\int^\infty_0dbb^2J_1(q_\perp b)\sum_{q,i,j}
e_q^2\Delta C^T_{q\leftarrow i}\otimes T_{i,F}(x_p,x_p;\mu_b)C_{\bar q\leftarrow j}\otimes f_{1,j/\pi}(x_\pi,\mu_b)
e^{-\left(S^{\mathrm{Siv}}_{\mathrm{NP}}+S^{f_{1q/\pi}}_{\mathrm{NP}}+S_\mathrm{P}\right)}.
\end{align}
Combing Eqs.~(\ref{eq:asy_Sivers}), (\ref{eq:dsig_UU_final}), (\ref{eq:dsig_UT_final}), one can get the Sivers asymmetry in the Drell-Yan process with a $\pi$ beam colliding on a transversely polarized proton target.

\subsection{The $\cos 2\phi$ asymmetry in the unpolarized Drell-Yan from double Boer-Mulders effect}

The angular differential cross section for unpolarized Drell-Yan process has the following general form
\begin{align}
\frac{1}{\sigma }\frac{d\sigma }{d\Omega }=\frac{3}{4\pi }\frac{1}{\lambda +3}(1+\lambda\cos^2\theta +\mu \sin2\theta \cos\phi+\frac{\nu }{2} \sin^2\theta \cos2\phi),
\label{eq:cross section}
\end{align}
where $\theta$ is the polar angle, and $\phi$ is the azimuthal angle of the hadron plane with respect to the dilepton plane in the Collins-Soper~(CS) frame~\cite{Collins:1977iv}.
The coefficients $\lambda$, $\mu$, $\nu$ in Eq.~(\ref{eq:cross section}) describe the sizes of different angular dependencies.
Particularly, $\nu$ stands for the asymmetry of the $\cos 2\phi$ azimuthal angular distribution of the dilepton.

The coefficients $\lambda$, $\mu$, $\nu$ have been measured in the process $\pi^-\, N\rightarrow \mu^+\mu^- \,X$  by the NA10 Collaboration~\cite{Falciano:1986wk,Guanziroli:1987rp} and the E615 Collaboration~\cite{Conway:1989fs} for a $\pi^-$ beam
with energies of 140, 194, 286 GeV, and 252 GeV,
with $N$ denoting a nucleon in the deuterium or tungsten target.
The experimental data showed a large value of $\nu$, near 30\% in the region $Q_T\sim3$ GeV.
This demonstrates a clear violation of the Lam-Tung relation~\cite{Lam:1978pu}.
In the last decade the angular coefficients were also measured in the $p\, N$ Drell-Yan processes in both the fixed target mode~\cite{Zhu:2006gx,Zhu:2008sj} and collider mode~\cite{Aaltonen:2011nr,Khachatryan:2015paa}.
The origin of large $\cos2\phi$ asymmetry--or the violation of the Lam-Tung relation-- observed in Drell-Yan process has been studied extensively in literature~\cite{Collins:1978yt,Chiappetta:1986yg,bran93,bran94,Eskola,Boer:1999mm,Boer:2006eq,Blazek,
Zhou:2009rp,Peng:2015spa,Lu:2016pdp,Lambertsen:2016wgj,Chang:2018pvk}.
Here we will only consider the contribution from the coupling of two Boer-Mulders functions from each hadron, denoted by $\nu_{\textrm{BM}}$.
It might be measured through the combination $2\nu_{\textrm{BM}}\approx 2\nu+\lambda-1$, in which the perturbative contribution is largely subtracted.

The $\cos 2\phi$ asymmetry coefficient $\nu_{\textrm{BM}}$ contributed by the Boer-Mulders function can be written as
\begin{align}
\nu_{\textrm{BM}}&=\frac{2\sum_q\mathcal{F}\left[\left(2\bm{\hat{h}}\cdot \bm{k}_\perp \bm{\hat{h}}\cdot \bm{p}_\perp-\bm{k}_{\perp}\cdot \bm{p}_\perp\right)\frac{h^\perp _{1,q/\pi}h^\perp_{1,\bar{q}/p}}{M_\pi M_p}\right]}{\sum_q\mathcal{F}\left[f_{1,q/\pi}f_{1,\bar{q}/p}\right]
} ,
\label{eq:nu}
\end{align}
where the notation
\begin{align}
\mathcal{F}[\omega f\bar{f}]= e_q^2\int {d}^{2}\bm{k}_\perp{d}^{2}\bm{p}_\perp{\delta
}^{2}(\bm{k}_\perp+\bm{p}_\perp-\bm{{q}}_{\perp})\omega f({x}_\pi,\bm{{k}}_\perp^2)\bar{f}({x}_p,\bm{p}_\perp^2)
\label{eq:notation}
\end{align}
has been adopted to express the convolution of transverse momenta.
$\bm{q}_\perp$, $\bm{k}_\perp$ and $\bm{p}_\perp$ are the transverse momenta of the lepton pair, quark and antiquark in the initial hadrons, respectively. $\bm{\hat{h}}$ is a unit vector defined as $\bm{\hat{h}}=\frac{\bm{q}_{\perp}}{|\bm{q}_{\perp}|} = \frac{\bm{q}_{\perp}}{q_\perp}$.
One can perform the Fourier transformation from $\bm{q}_\perp$ space to $\bm{b}$ space on the delta function in the notation of Eq.~(\ref{eq:notation}) to obtain the denominator in Eq.~(\ref{eq:nu}) as
\begin{align}
\mathcal{F}\left[f_{1,q/\pi}f_{1,\bar{q}/p}\right]&=\sum_q e_q^2\int \frac{d^2b}{(2\pi)^2} \int d^2\bm{k}_\perp d^2\bm{p}_\perp e^{i(\bm{{q}}_{T}-\bm{k}_\perp-\bm{p}_\perp)\cdot \bm{b}}f_{1,q/\pi}({x}_\pi,\bm{k}_\perp^2)f_{1,\bar{q}/p}({x}_p,\bm{p}_\perp^2)\nonumber\\
&=\sum_q e_q^2\int \frac{d^2b}{(2\pi)^2} e^{i\bm{q}_{\perp}\cdot \bm{b}} \widetilde{f}_{1,q/\pi}(x_\pi,b;Q)\widetilde{f}_{1,\bar{q}/p}(x_p,b;Q)
\label{eq:denominator}
\end{align}
where the unpolarized distribution function in $b$ space is given in Eqs.~(\ref{eq:unpo_f}) and (\ref{eq:unpo_fpi}).
Similar to the treatment of the denominator, using the expression of the Boer-Mulders function in Eq.~(\ref{eq:BM_b}) the numerator can be obtained as
\begin{align}
&\mathcal{F}\left[\left(2\bm{\hat{h}}\cdot \bm{k}_\perp \bm{\hat{h}}\cdot \bm{p}_\perp-\bm{k}_{\perp}\cdot \bm{p}_\perp\right)\frac{h^{\perp}_{1,q/\pi}h^{\perp}_{1,\bar{q}/p}}{M_\pi M_p}\right]\nonumber\\
&=\sum_q e_q^2\int \frac{d^2b}{(2\pi)^2} \int d^2\bm{k}_\perp d^2\bm{p}_\perp e^{i(\bm{{q}}_{T}-\bm{k}_\perp-\bm{p}_\perp)\cdot b}\left[\left(2\bm{\hat{h}}\cdot \bm{k}_\perp \bm{\hat{h}}\cdot \bm{p}_\perp-\bm{k}_{\perp}\cdot \bm{p}_\perp\right)\frac{h^{\perp}_{1,q/\pi}h^{\perp}_{1,\bar{q}/p}}{M_\pi M_p}\right] \nonumber\\
&=\sum_q e_q^2\int \frac{d^2b}{(2\pi)^2} e^{i\bm{q}_\perp\cdot \bm{b}} (2\hat{h}_\alpha\hat{h}_\beta-g_{\alpha\beta}^\perp)\widetilde{h}^{\alpha\perp }_{1,q/\pi}(x_\pi,b;Q)\widetilde{h}^{\beta\perp}_{1,\bar{q}/p}(x_p,b;Q)\nonumber \\
&=\sum_q e_q^2\int_0^\infty \frac{dbb^3} {8\pi}J_2(q_\perp b)T^{(\sigma)}_{q/\pi,F}(x_\pi,x_\pi;\mu_b)
T^{(\sigma)}_{\bar{q}/p,F}(x_p,x_p;\mu_b)
e^{-\left(S^{f_{1,q/p}}_{\mathrm{NP}}+S^{f_{1,q/\pi}}_{\mathrm{NP}}+S_\mathrm{P}\right)}.
\label{eq:numerator}
\end{align}
Different from the previous two cases, the hard coefficients $C^{\rm BM}_i$ and $\mathcal{H}_{\rm BM}$ for the Boer-Mulders function have not been calculated up to next-to-leading order (NLO), and still remain in leading order (LO) as $C_{q\leftarrow i}=\delta_{qi}\delta(1-x)$ and $\mathcal{H} =1$.

\section{Numerical estimate for the TMD distributions}
\label{Sec.TMDs_num}

Based on the TMD evolution formalism for the distributions set up in Sec.~\ref{Sec.TMDs_theo}, we will show the numerical results for the TMD distributions. Particularly attention will be paid on those of the pion meson, as those of the proton has the been studied numerically in Refs.~\cite{Aybat:2011zv,Echevarria:2014rua,Su:2014wpa}.

\subsection{The unpolarized TMD distribution of the pion meson}

\begin{figure}
\centering
\includegraphics[width=0.8\columnwidth]{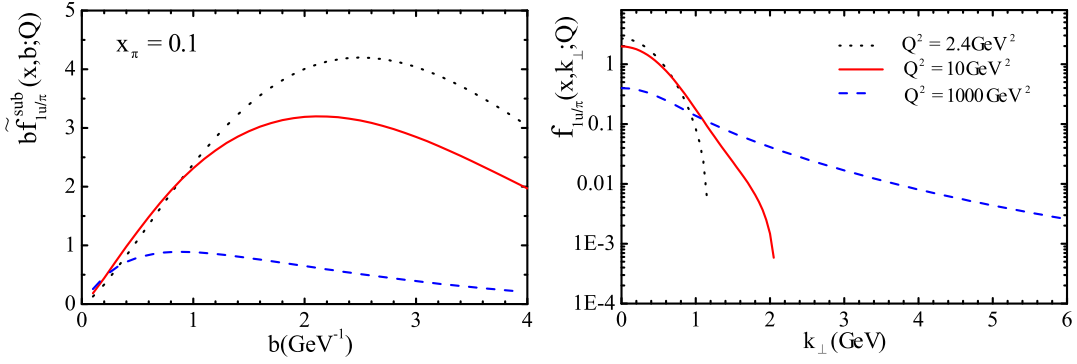}
\caption{Subtracted unpolarized TMD distribution of the pion meson for valence quarks in $b$-space (left panel) and $k_\perp$-space (right panel), at energies: $Q^2=2.4\ \mathrm{GeV}^2$~(dotted lines), $Q^2=10\ \mathrm{GeV}^2$~(solid lines) and $Q^2=1000\ \mathrm{GeV}^2$~(dashed lines). From Ref.~\cite{Wang:2017zym}.}
\label{fig:fsub_pion}
\end{figure}

In Ref.~\cite{Wang:2017zym}, the authors applied Eq.~(\ref{eq:unpo_fpi}) and the extracted parameters $g_1^\pi$ and $g_2^\pi$ to quantitatively study the scale dependence of the unpolarized TMD distributions of the pion meson with the JCC Scheme.
For the collinear unpolarized distribution function of the pion meson, the
NLO SMRS parametrization~\cite{Sutton:1991ay} was chosen.
The results are plotted in Fig.~\ref{fig:fsub_pion}, with the left and right panels showing the subtracted distribution in $b$ space  and $k_\perp$ space, for fixed $x_\pi=0.1$, at three different energy scales: $Q^2=2.4\ \mathrm{GeV}^2$ (dotted line), $10\ \mathrm{GeV}^2$ (solid line), $1000\ \mathrm{GeV}^2$ (dashed line).
From the $b$-dependent plots, one can see that at the highest energy scale $Q^2=1000\ \mathrm{GeV}^2$,  the peak of the curve is in the low $b$ region where $b<b_{\mathrm{max}}$, since in this case the perturbative part of the Sudakov form factor dominates.
However, at lower energy scales, e.g., $Q^2=10\ \mathrm{GeV}^2$ and $Q^2=2.4\ \mathrm{GeV}^2$, the peak of the $b$-dependent distribution function moves towards the higher $b$ region, indicating that the nonperturbative part of the TMD evolution becomes important at lower energy scales.
For the distribution in $k_\perp$ space, at higher energy scale the distribution has a tail falling off slowly at large $k_\perp$, while at lower energy scales the distribution function falls off rapidly with increasing $k_\perp$.
It is interesting to point out that the shapes of the pion TMD distribution at different scales are similar to those of the proton, namely, Fig.~8 in Ref.~\cite{Su:2014wpa}.

\subsection{The Sivers function of the proton}

The scale dependence of the T-odd distributions, such as the Sivers function and the Boer-Mulders function, is more involved than that of the T-even distributions.
This is because their collinear counterparts are the twist-3 multiparton correlation functions~\cite{Echevarria:2014xaa,Kang:2011mr,Sun:2013hua,Wang:2018naw,Wang:2018pmx}, for which
the exact evolution equations are far more complicated than those for the unpolarized distribution function. In numerical calculation, some approximations on the evolution kernels are usually adopted.

\begin{figure}
\centering
\includegraphics[width=0.48\columnwidth]{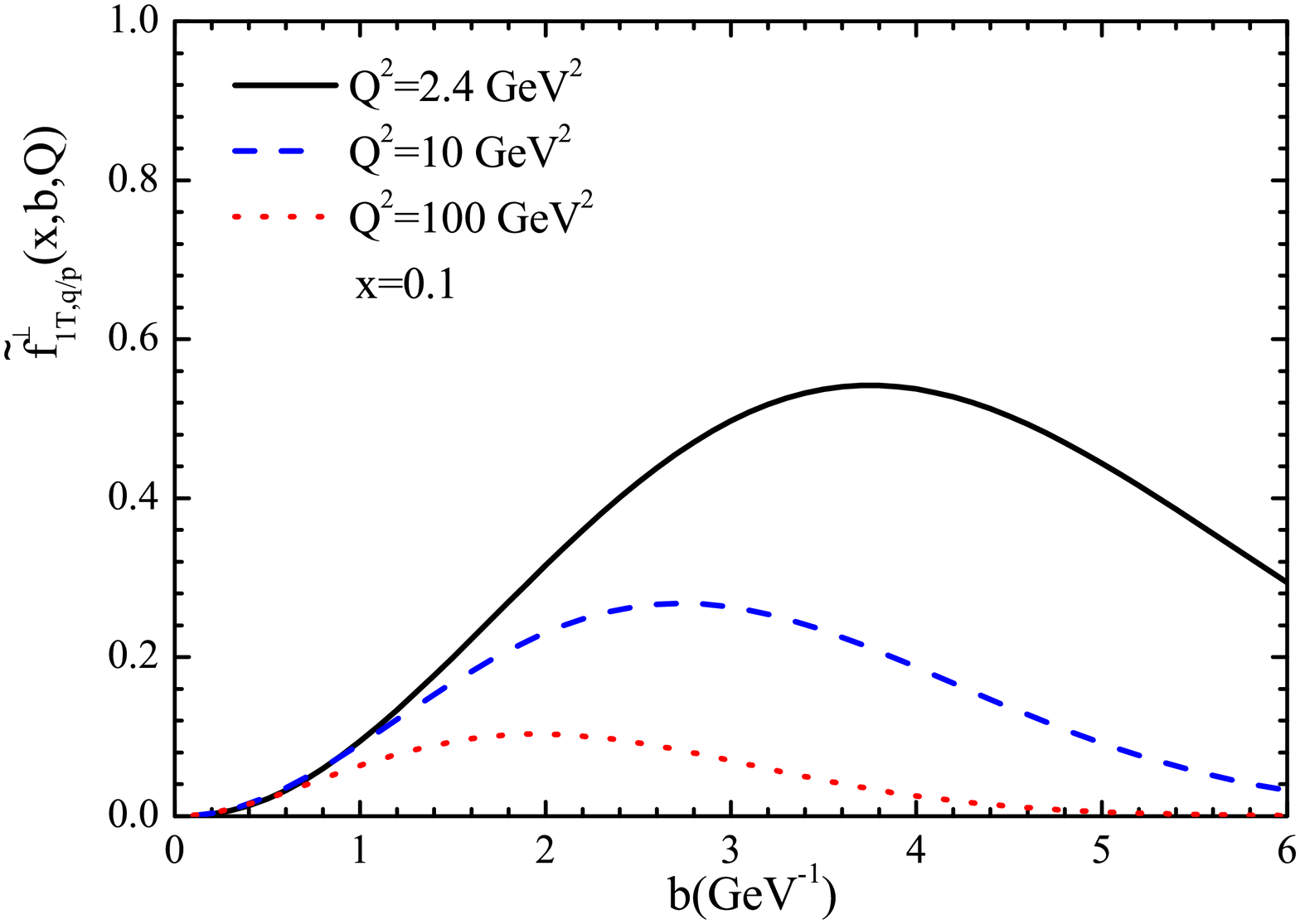}
\includegraphics[width=0.48\columnwidth]{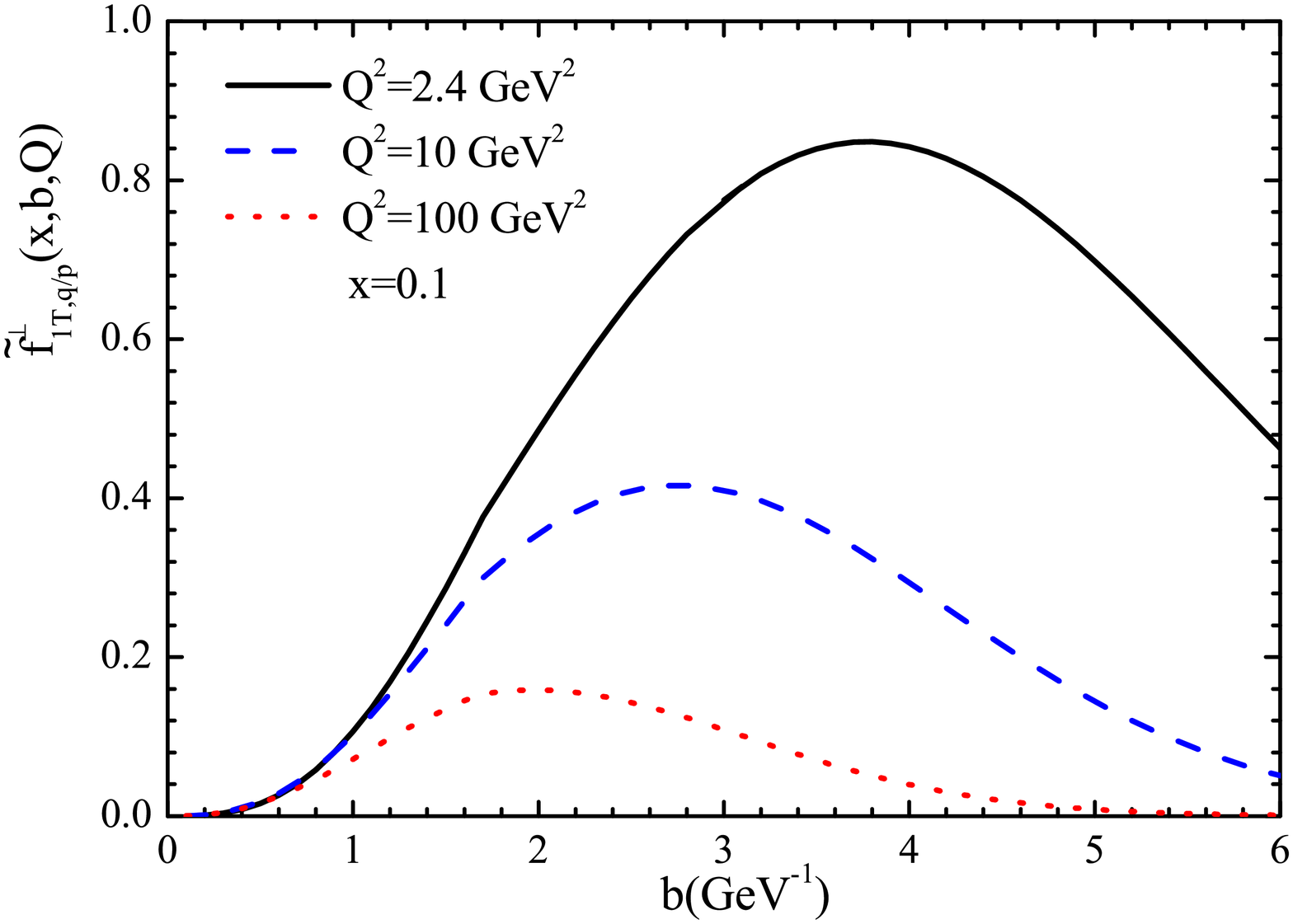}\\
\includegraphics[width=0.48\columnwidth]{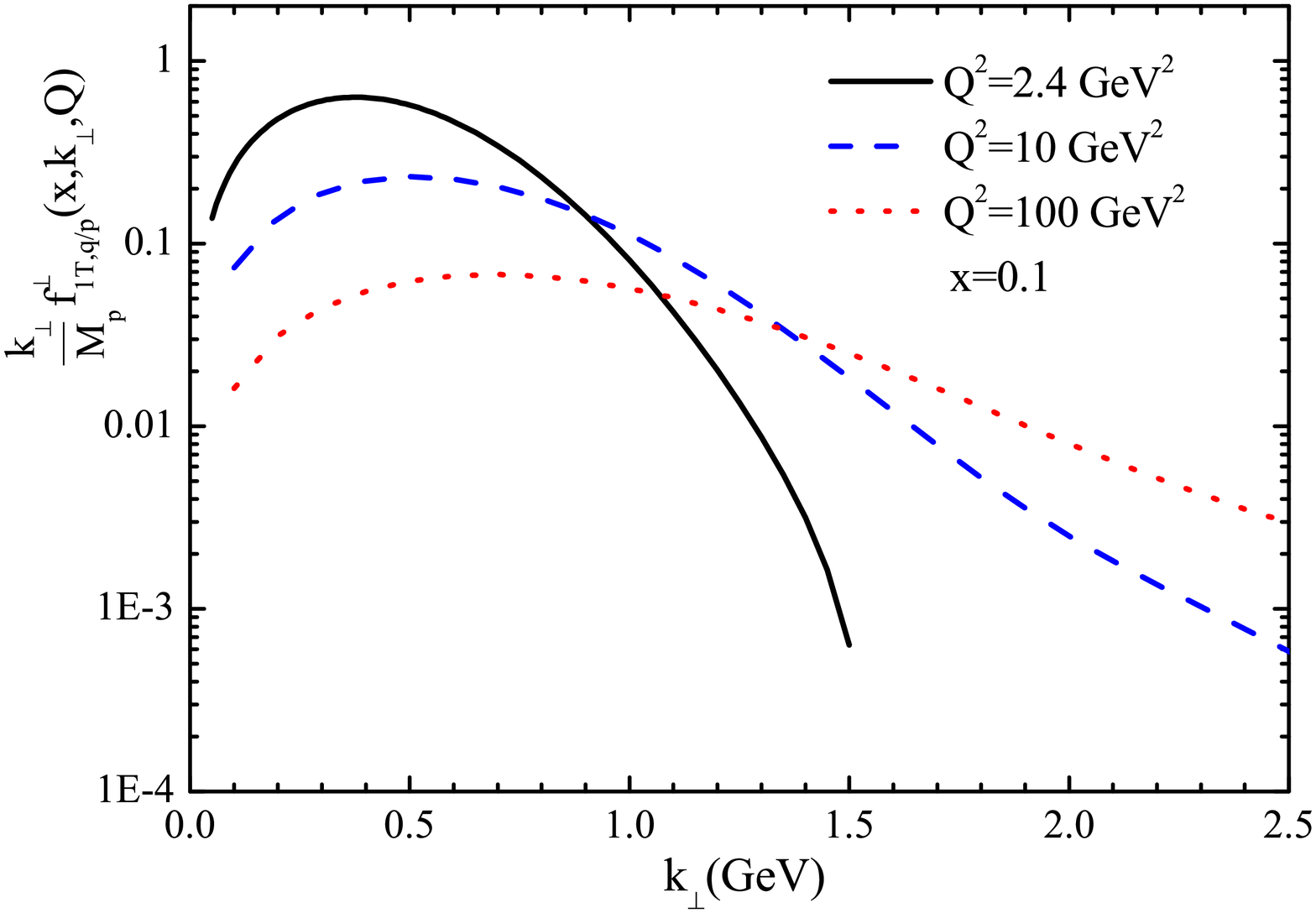}
\includegraphics[width=0.48\columnwidth]{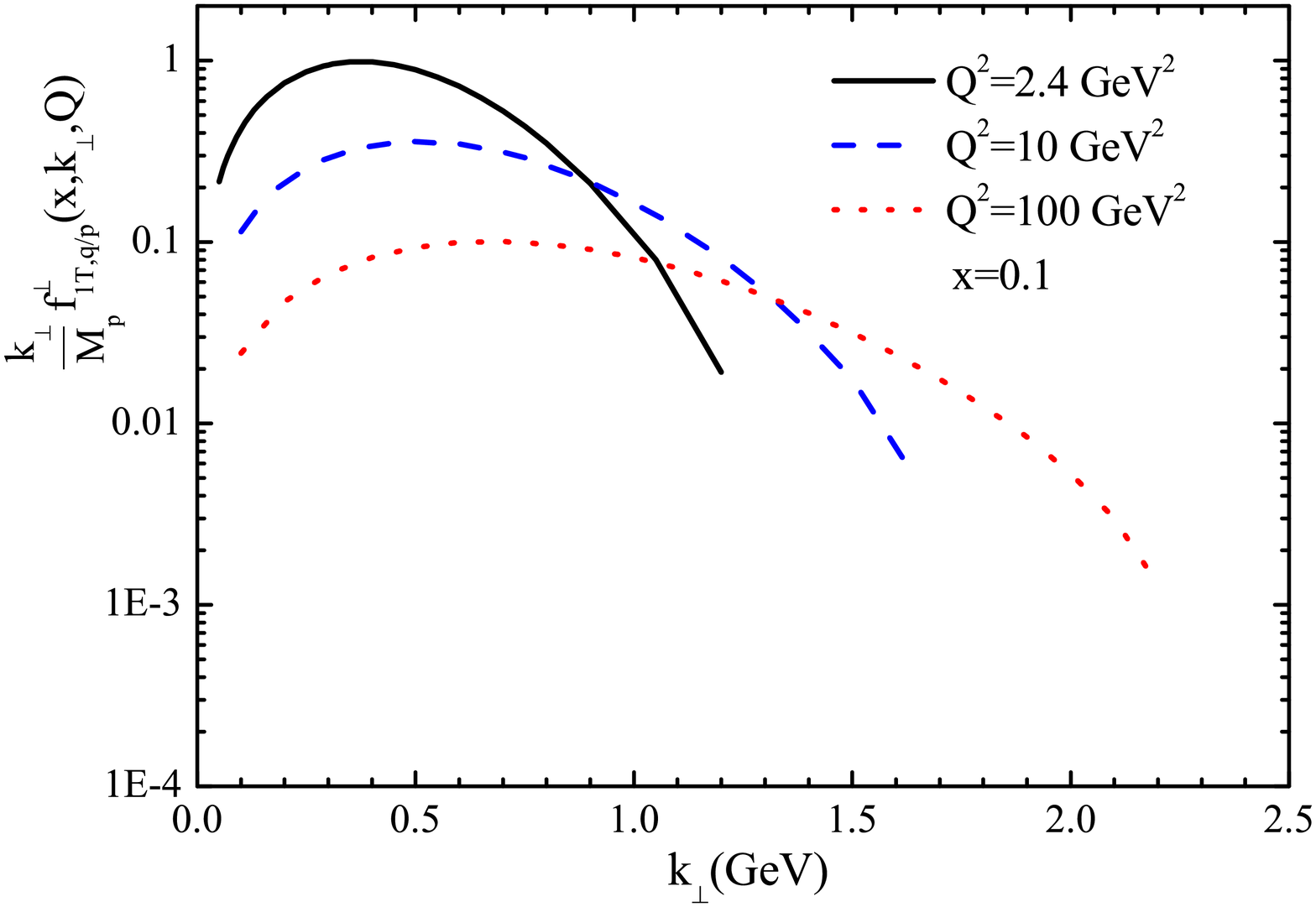}
\caption{Subtracted Sivers function for the up quarks in Drell-Yan in $b$-space (upper panels) and $k_\perp$-space (lower panels), at energies: $Q^2=2.4\ \mathrm{GeV}^2$~(solid lines), $Q^2=10\ \mathrm{GeV}^2$~(dashed lines) and $Q^2=100\ \mathrm{GeV}^2$~(dotted lines). The left panel shows the result from a same evolution kernel as that for $f_1$ in Eq.~(\ref{eq:dglap}), the right panel shows the result from an approximate evolution kernel for the Qiu-Sterman function in Eq.~(\ref{eq:qs}), respectively. Figure from Ref.~\cite{Wang:2018pmx}}
\label{fig:Sivers_fun}
\end{figure}

In Ref.~\cite{Echevarria:2014xaa}, the Qiu-Sterman function was assumed to be proportional to $f_1$, namely, it follows the same evolution kernel as that for $f_1$.
A different choice was adopted in Ref.~\cite{Wang:2018pmx}, where the homogenous terms of the exact evolution kernel for the Qiu-Sterman function~\cite{Kang:2012em,Kang:2008ey,Zhou:2008mz,Vogelsang:2009pj,
Braun:2009mi,Ma:2011nd,Schafer:2012ra,Ma:2012xn,Sun:2013hua,Zhou:2015lxa} were included to deal with the scale dependence of Qiu-Sterman function:
\begin{align}
P^{\mathrm{QS}}_{qq}\approx P^{f_1}_{qq}-\frac{N_c}{2}\frac{1+z^2}{1-z}-N_c\delta(1-z),
\label{eq:qs}
\end{align}
with $P^{f_1}_{qq}$ the evolution kernel of the unpolarized PDF
\begin{align}
P^{f_1}_{qq}=\frac{4}{3}\left(\frac{1+z^2}{(1-z)_+}+\frac{3}{2}\delta(1-z)\right).\label{eq:dglap}
\end{align}
To solve the QCD evolution numerically, we resort to the QCD evolution package HOPPET~\cite{Salam:2008qg} and we custom the code to include the splitting function in Eq.~(\ref{eq:qs}).
For a comparison, in Fig.~\ref{fig:Sivers_fun} we plot the TMD evolution of the Sivers function for proton in $b$ space and the $k_\perp$ space using the above mentioned two approaches~\cite{Wang:2018pmx}.
In this estimate, the next leading order $C$-coefficients $\Delta C^T_{q\leftarrow i}$ was adopted from Ref.~\cite{Kang:2011mr,Sun:2013hua} and the nonperturbative Sudakov form factor for the Sivers function of proton was adopted as the form in Eq.~(\ref{snp:gaussian}).
The Sivers functions are presented at three different energy scales: $Q^2=2.4 \,\mathrm{GeV}^2$, $Q^2=10 \, \mathrm{GeV}^2$ and $Q^2=100\, \mathrm{GeV}^2$.
Similar to the result for $f_1$, one can conclude from the curves that the perturbative Sudakov form factor dominated in the low $b$ region at higher energy scales and the nonperturbative part of the TMD evolution became more important at lower energy scales. However, the $k_\perp$ tendency of the Sivers function in the two approaches is different, which indicates that the scale dependence of the Qiu-Sterman function may play a role in the TMD evolution.

\subsection{The Boer-Mulders function of the pion meson}
\label{sec:bm_pion}

The evolution of the Boer-Mulders function for the valence quark inside $\pi$ meson has been calculated from  Eqs.~(\ref{eq:BM_b}) and (\ref{eq:BM_kt}) in Ref.~\cite{Wang:2018naw}, in which
the collinear twist-3 correlation function $T_{q,F}^{(\sigma)}$ at the initial energy scale was obtained by adopting a model result of the Boer-Mulders function of the pion meson calculated from the light-cone wave functions~\cite{Wang:2017onm}.
For the scale evolution of $T^{(\sigma)}_{q,F}$, the exact evolution effect has been studied in Ref.~\cite{Kang:2012em}.
For our purpose, we only consider the homogenous term in the evolution kernel
\begin{align}
P^{T^{(\sigma)}_{q,F}}_{qq}(x)\approx\Delta_T\,P_{qq}(x)-N_C\delta(1-x),\label{eq:evobm}
\end{align}
with $\Delta_T\,P_{qq}(x)=C_F\left[\frac{2z}{(1-z)_+}+\frac{3}{2}\delta(1-x)\right]$ being the evolution kernel for the transversity distribution function $h_1(x)$. We customize the original code of {\sc{QCDNUM}}~\cite{Botje:2010ay} to include the approximate kernel in Eq.~(\ref{eq:evobm}). For the nonperturbative part of the Sudakov form factor associated with Boer-Mulders function,
the information still remains unknown.
The assumption that $S_{\rm NP}$ for Boer-Mulders function is same as that for $f_1$ can be a practical way to access the information of TMD evolution for Boer-Mulders function.

\begin{figure}
  \centering
  \includegraphics[width=0.48\columnwidth]{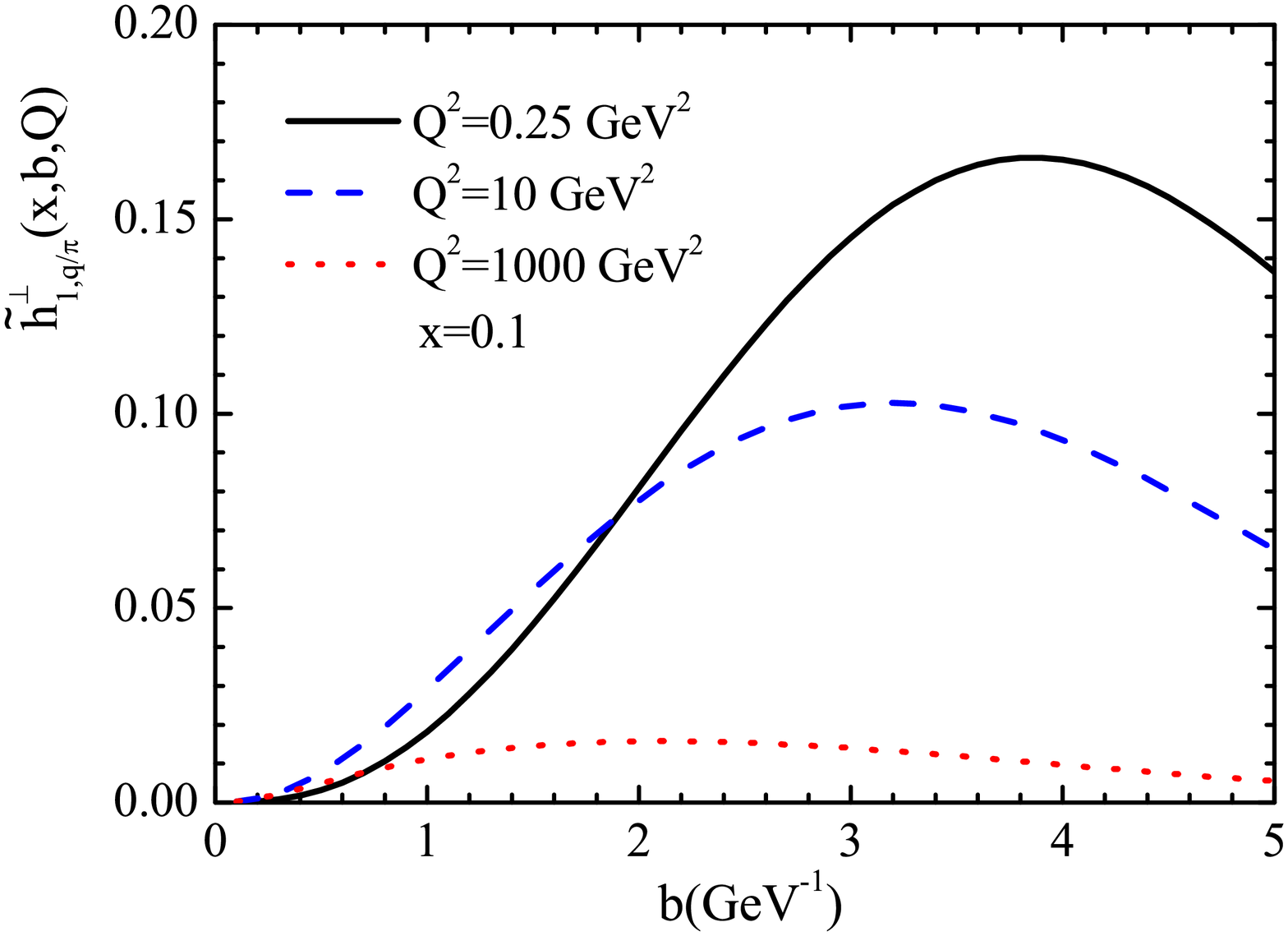}
  \includegraphics[width=0.48\columnwidth]{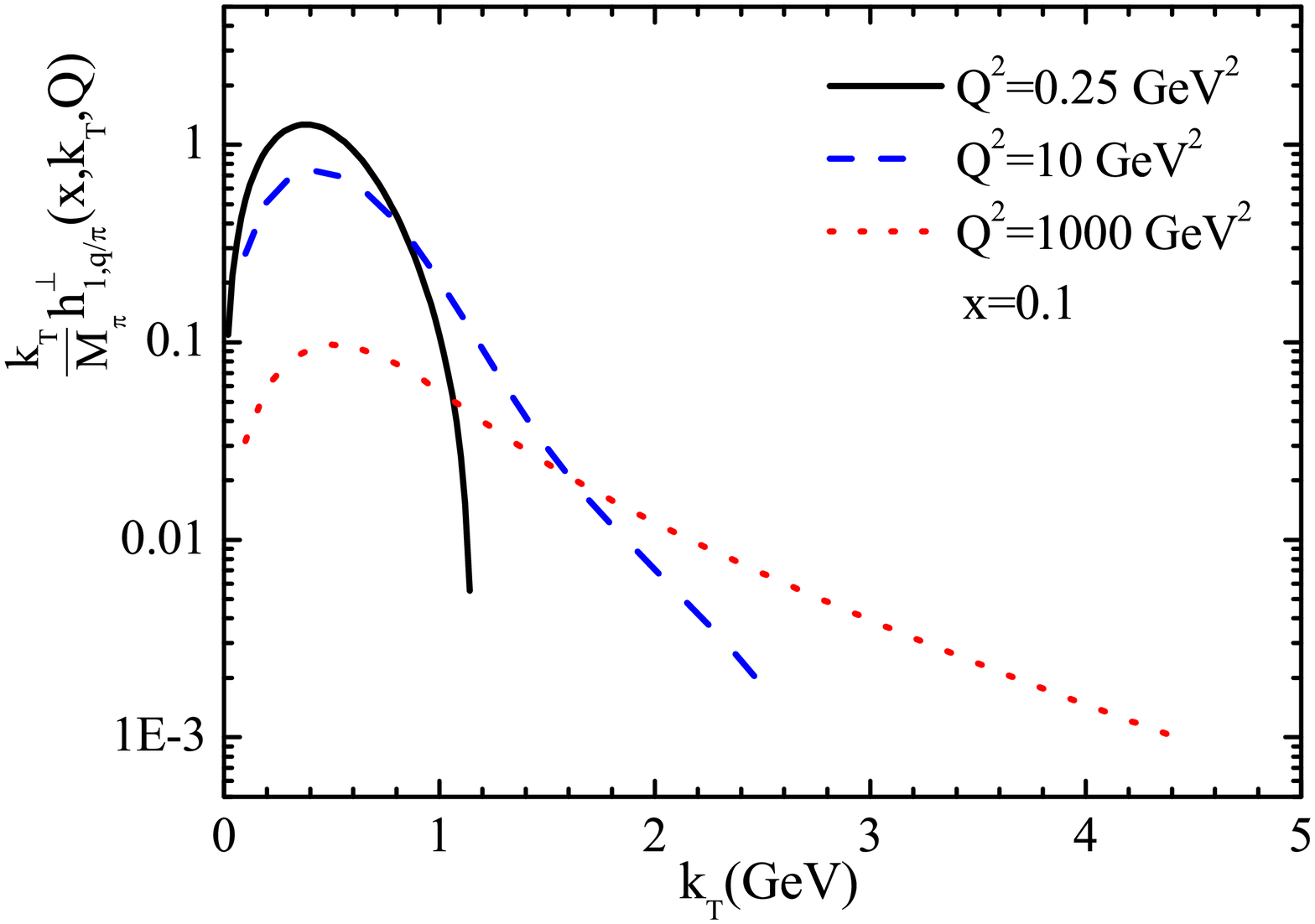}\\
  \caption{The Boer-Mulders function for $u$-quark in $b$ space~(left panel) and $k_T$ space~(right panel) considering three different energy scales: $Q^2=2.4 \mathrm{GeV}^2$~(solid lines), $Q^2=10 \mathrm{GeV}^2$~(dashed lines), $Q^2=1000 \mathrm{GeV}^2$~(dotted lines). Figures from Ref.~\cite{Wang:2018naw}.}
  \label{fig:BM}
\end{figure}

We plot the $b$-dependent and $k_T$-dependent Boer-Mulders function at $x=0.1$ in the left and right panels of Fig.~\ref{fig:BM}, respectively.
In calculating $\tilde{h}_{1,q/\pi}^{\perp}(x,b;Q)$ in Fig.~\ref{fig:BM}, we have
rewritten the Boer-Mulders function in $b$ space as
\begin{align}
&\tilde{h}_{1,q/\pi}^{\perp}(x,b;Q)= \frac{i b_\alpha}{\pi} \tilde{h}_{1,q/\pi}^{\alpha\perp}(x,b;Q).
\end{align}
The three curves in each panel correspond to three different energy scales: $Q^2=0.25 \mathrm{GeV}^2$~(solid lines), $Q^2=10 \mathrm{GeV}^2$~(dashed lines), $Q^2=1000 \mathrm{GeV}^2$~(dotted lines).
From the curves, we find that the TMD evolution effect of the Boer-Mulders function is significant and should  be considered in phenomenological analysis.
The result also indicates that the perturbative Sudakov form factor
dominates in the low $b$ region at higher energy scales and the nonperturbative part of the TMD evolution becomes more important at lower energy scales.

As a conclusion, we find that the tendency of the distributions is similar: the distribution is dominated by perturbative region in $b$ space at large $Q^2$, while at lower $Q^2$ the distribution shifts to the large $b$ region, indicating that the nonperturbative effects of TMD evolution become important. For the distributions in $k_\perp$ space, as the value of $Q^2$ increases, the distributions become wider with a perturbative tail, while at low values of $Q^2$, the distributions resemble gaussian-type parametrization.
However, the widths of the transverse momentum differ among different distributions.

\section{Numerical estimate for the physical observables in $\pi$-$N$ Drell-Yan process}
\label{Sec.DY_num}

Based on the general TMD factorization framework provided in Sec.~\ref{Sec.DY_theo}, we present several physical observables in $\pi$-$N$ Drell-Yan process in this section.

QCD predicts that the T-odd PDFs present generalized universality, i.e., the sign of the Sivers function measured in Drell-Yan process should be opposite to its sign measured in SIDIS~\cite{Collins:2002kn,Brodsky:2002rv,Brodsky:2002cx} process. The verification of this sign change~\cite{Anselmino:2009st,Kang:2009bp,Peng:2014hta,Echevarria:2014xaa,Huang:2015vpy,Anselmino:2016uie} is one of the most fundamental tests of our understanding of the QCD dynamics and the factorization scheme, and it is also the main pursue of the existing and future Drell-Yan facilities~\cite{Aghasyan:2017jop,Gautheron:2010wva,Fermilab1,Fermilab2,ANDY,Adamczyk:2015gyk}.
The COMPASS Collaboration has reported the first measurement of the Sivers asymmetry in the pion-induced Drell-Yan process, in which a $\pi^-$ beam was scattered off the transversely polarized NH$_3$ target~\cite{Aghasyan:2017jop}. The polarized Drell-Yan data from COMPASS, together with the previous measurement of the Sivers effect in the $W$- and $Z$-boson production from $p^\uparrow p$ collision at RHIC~\cite{Adamczyk:2015gyk} will provide the first evidence of the sign change of the Sivers function.
As COMPASS experiment has almost the same setup~\cite{Aghasyan:2017jop,Adolph:2016dvl} for SIDIS and Drell-Yan process, it will provide the unique chance to explore the sign change since the uncertainties in the extraction of the Sivers function from the two kinds of measurements can be reduced.

\subsection{The normalized cross section for unpolarized $\pi$-$N$ Drell-Yan process}

\begin{figure}
  \centering
  \includegraphics[width=0.5\columnwidth]{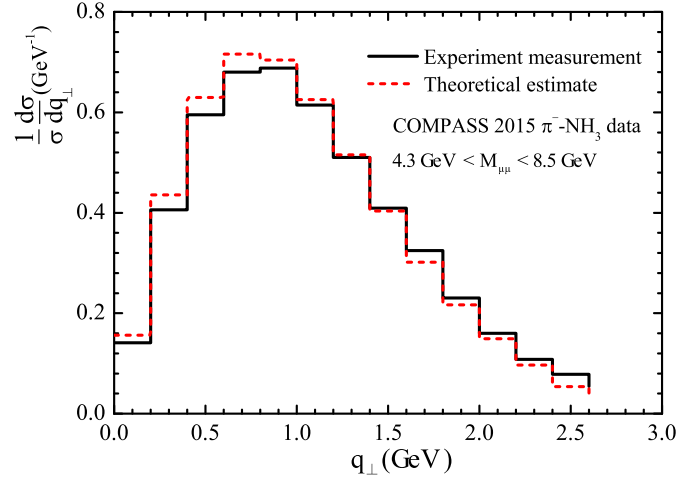}
  \caption{The transverse spectrum of lepton pair production in the unpolarized pion-nucleon Drell-Yan process, with an NH$_3$ target at COMPASS.
  The dashed line represents the theoretical calculation in Ref.~\cite{Wang:2017zym}.
  The solid line shows the experimental measurement at COMPASS~\cite{Aghasyan:2017jop}. Figure from   Ref.~\cite{Wang:2017zym}.}
  \label{fig:COMPASS}
\end{figure}

The very first step to understand the Sivers asymmetry in the $\pi$-$N$ Drell-Yan process is to quantitatively estimate the differential cross section in the same process for unpolarized nucleon target with high accuracy, since it always appears in the denominator of the asymmetry definition.
The differential cross section for unpolarized Drell-Yan process has been given in Eq.~(\ref{eq:dsig_UU_final}).
Applying the extracted nonperturbative Sudakov form factor for pion meson in Ref.~\cite{Wang:2017zym} and the extracted $S_{\rm NP}$ for nucleon in Ref.~\cite{Su:2014wpa}, we estimated the normalized transverse momentum spectrum of the dimuon production in the pion-nucleon Drell-Yan process at COMPASS for different $q_\perp$ bins with an interval of 0.2 GeV.
The result is plotted in Fig.~\ref{fig:COMPASS}.
From the curves, one can find the theoretical estimate on the $q_\perp$ distribution of the dimuon agreed with the COMPASS data fairly well in the small $q_\perp$ region where the TMD factorization is supposed to hold.
The comparison somehow confirms the validity of extraction of the nonperturbative Sudakov form factor for the unpolarized distribution $f_{1\pi}$ of pion meson, within the TMD factorization.
This may indicate that the framework can also be extended to the study of the azimuthal asymmetries in the $\pi N$ Drell-Yan process, such as the Sivers asymmetry and Boer-Mulders asymmetry.
We should point out that at larger $q_\perp$, the numerical estimate in Ref.~\cite{Wang:2017zym} cannot describe the data, indicating that the perturbative correction from the $Y_{UU}$ term may play an important role in the region $q_\perp \sim Q$.
Further study on the $Y$ term is needed to provide a complete picture of the $q_\perp$ distribution of lepton pairs from $\pi N$ Drell-Yan in the whole $q_\perp$ range.

\subsection{The Sivers asymmetry}

\begin{figure}
\centering
\includegraphics[width=0.48\columnwidth]{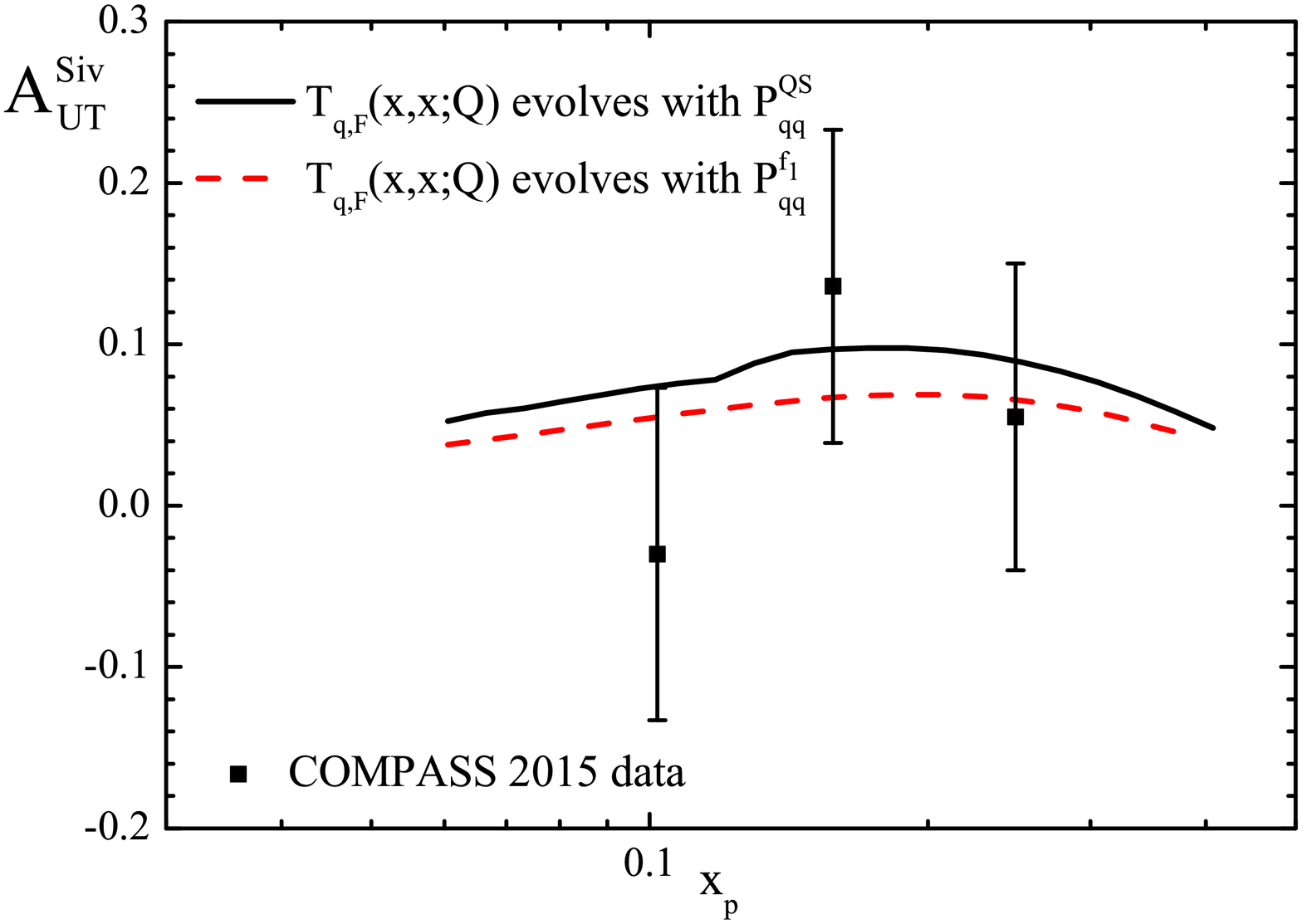}\quad
\includegraphics[width=0.48\columnwidth]{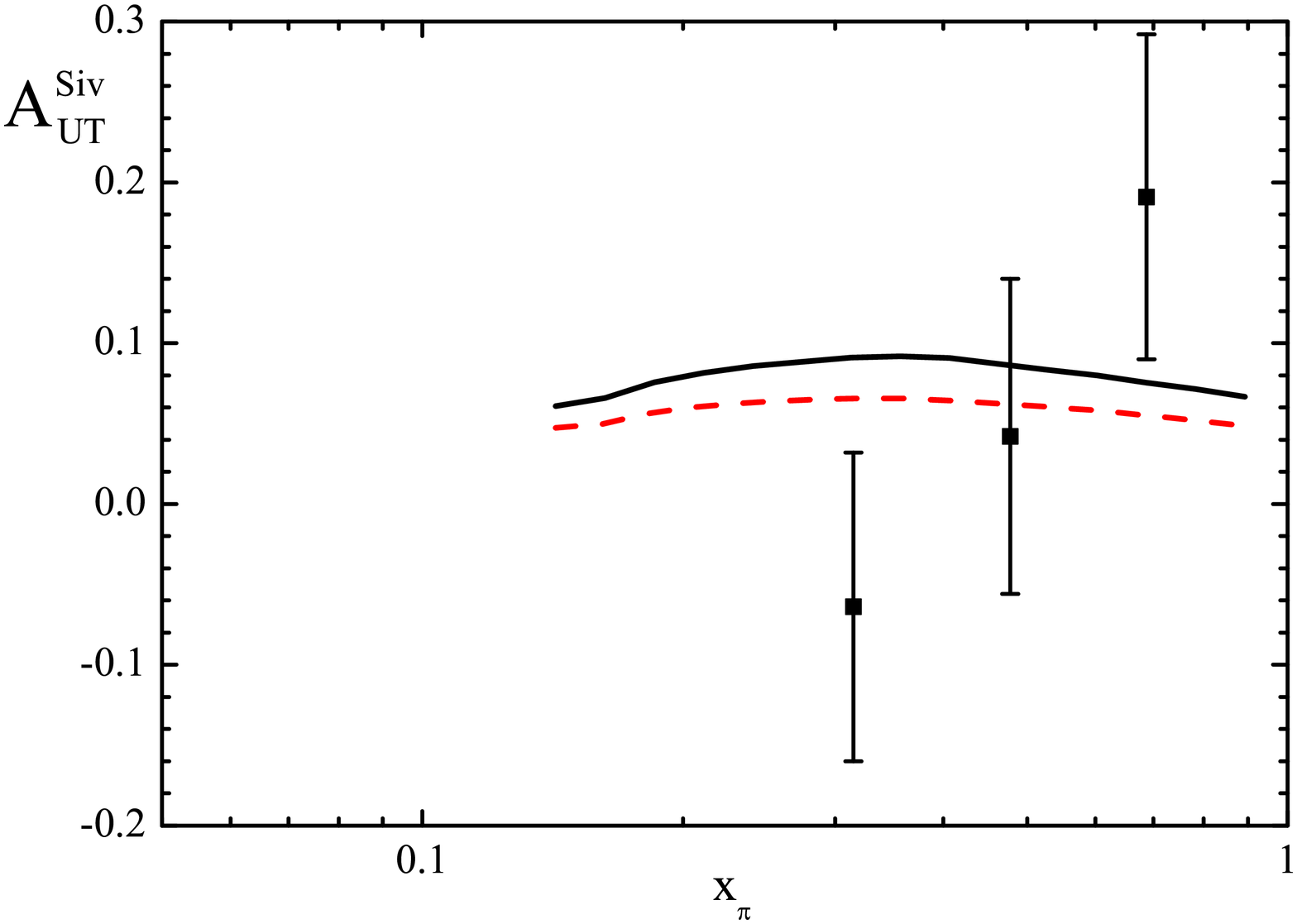}
\includegraphics[width=0.48\columnwidth]{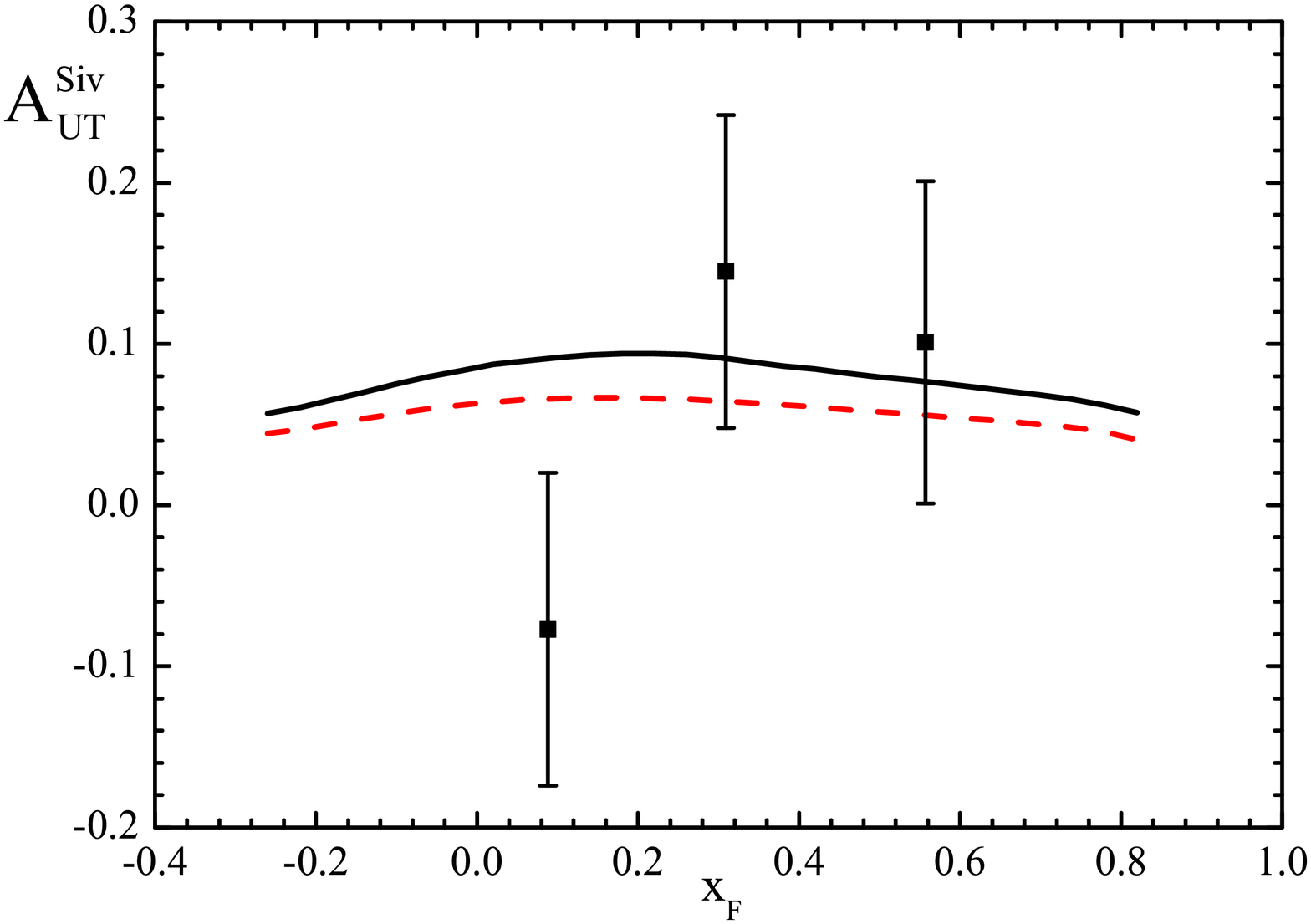}\quad
\includegraphics[width=0.48\columnwidth]{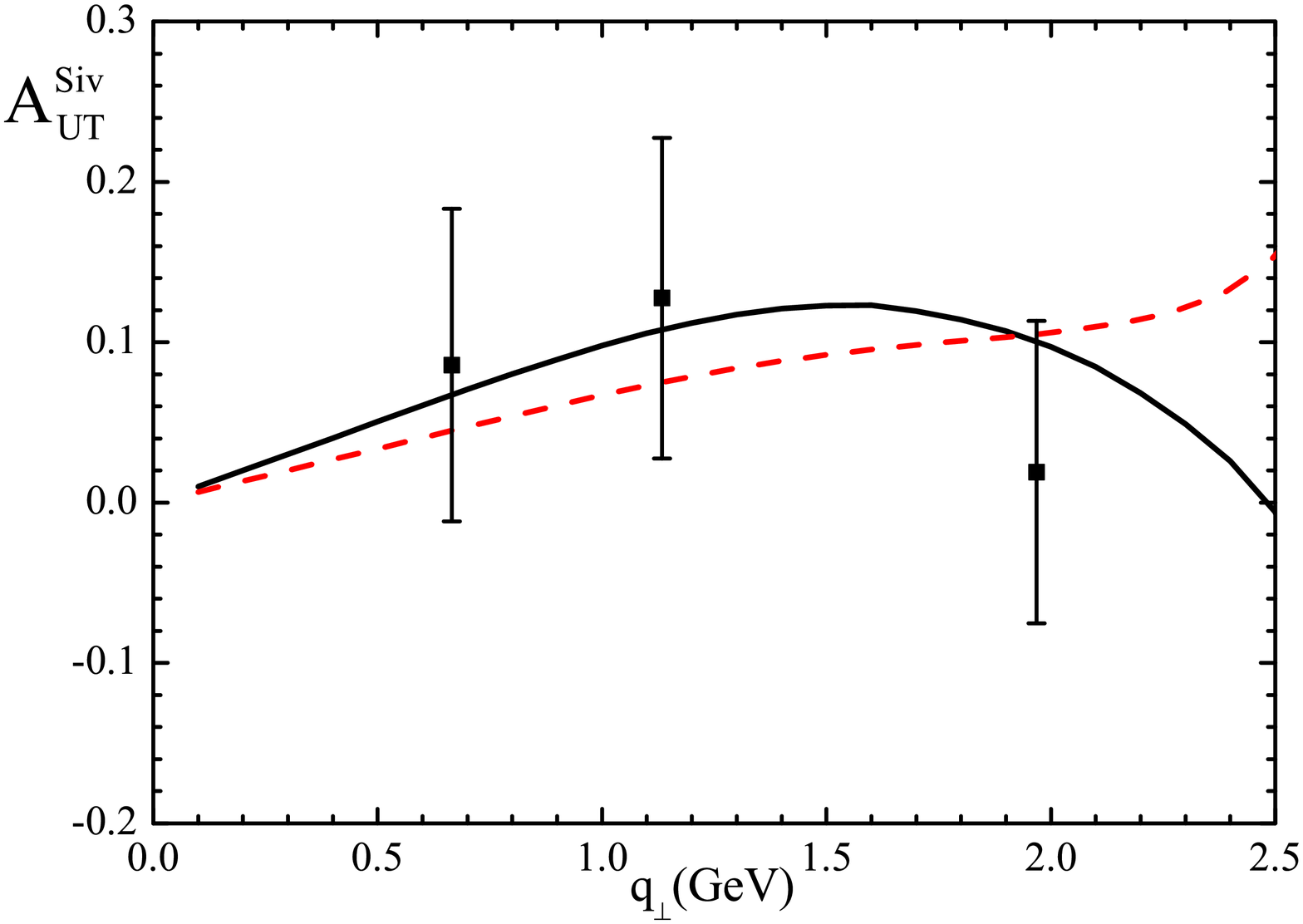}
\caption{The Sivers asymmetry within TMD factorization for a $\pi^-$ scattering off transversely polarized proton Drell-Yan process as functions of $x_p$~(upper left), $x_\pi$~(upper right), $x_F$~(lower left) and $q_\perp$~(lower right), compared with the COMPASS data~\cite{Aghasyan:2017jop}. Figure from Ref.~\cite{Wang:2018pmx}.}
\label{fig:Sivers_asy}
\end{figure}

In Ref.~\cite{Echevarria:2014xaa}, the authors adopted the Gaussian form of the nonperturbative Sudakov form factor $S_{\rm NP}$ in Eq.~(\ref{snp:gaussian}) and the leading order $C$ coefficients to perform a global fit on the Sivers function from the experimental data at HERMES~\cite{Airapetian:2009ae},
COMPASS~\cite{Alekseev:2008aa,Adolph:2012sp}, and Jefferson Lab (JLab)~\cite{Qian:2011py}.
With the extracted Sivers function from SIDIS process at hand, they made predictions for the Sivers asymmetry in Drell-Yan lepton pair and $W$ production at future planned Drell-Yan facilities at COMPASS~\cite{Gautheron:2010wva}, Fermilab~\cite{Fermilab1,Fermilab2} and RHIC~\cite{Aschenauer:2013woa,ANDY}, which can be compared to the future experimental measurements to
test the sign change of the Sivers functions between SIDIS and Drell-Yan processes. The predictions were presented in Fig.~12 and 13 of Ref.~\cite{Echevarria:2014xaa}.

The TMD evolution effect of the Sivers asymmetry in SIDIS and $pp$ Drell-Yan at low transverse momentum has also been studied in Ref.~\cite{Sun:2013hua}, in which a framework was build to match SIDIS
and Drell-Yan and cover the TMD physics with $Q^2$ from several $\textmd{GeV}^2$ to $10^4\textmd{GeV}^2$ (for $W/Z$ boson production). It has shown that the evolution equations derived by a direct integral of the CSS evolution kernel from low to high $Q$ can describe well the transverse momentum distribution of the unpolarized cross sections in the $Q^2$ range from 2 to 100 GeV$^2$.
With this approach, the transverse moment of the quark Sivers functions can be constrained from the combined analysis of the HERMES and COMPASS data on the Sivers asymmetries in SIDIS.
Based on this result, Ref.~\cite{Sun:2013hua} provided the predictions for the Sivers asymmetries
in $pp$ Drell-Yan, as well as in $\pi^- p$ Drell-Yan.
The latter one has been measured by the COMPASS collaboration, and the comparison showed that the theoretical result is consistent with data (Fig.~6 in Ref.~\cite{Aghasyan:2017jop}) within the error bar.

With the numerical results of the TMD distributions in Eq.~(\ref{eq:WUT}), the Sivers asymmetry $A_{UT}^{\textrm{Siv}}$ as function of $x_p$, $x_\pi$, $x_F$ and $q_\perp$  in $\pi^- p^\uparrow \to \mu^+ \mu^- +X$ in the kinematics of COMPASS Collaboration was calculated in Ref.~\cite{Wang:2018pmx}, as shown in Fig.~\ref{fig:Sivers_asy}.
The magnitude of the asymmetry is around $0.05 \div 0.10$, which is consistent with the COMPASS measurement (full squares in Fig.~\ref{fig:Sivers_asy})~\cite{Aghasyan:2017jop} within the uncertainties of the asymmetry.
The different approaches dealing with the energy dependence of Qiu-Sterman function lead to different shapes of the asymmetry.
Furthermore, the asymmetry from the approximate evolution kernel has a fall at larger $q_\perp$, which is more compatible to the shape of $q_\perp$-dependent asymmetry of measured by the COMPASS Collaboration.
The study may indicate that, besides the TMD evolution effect, the scale dependence of the Qiu-Sterman function will also play a role in the interpretation of the experimental data.

\subsection{The $\cos 2\phi$ azimuthal asymmetry}

\begin{figure}
  \centering
  \includegraphics[width=0.48\columnwidth]{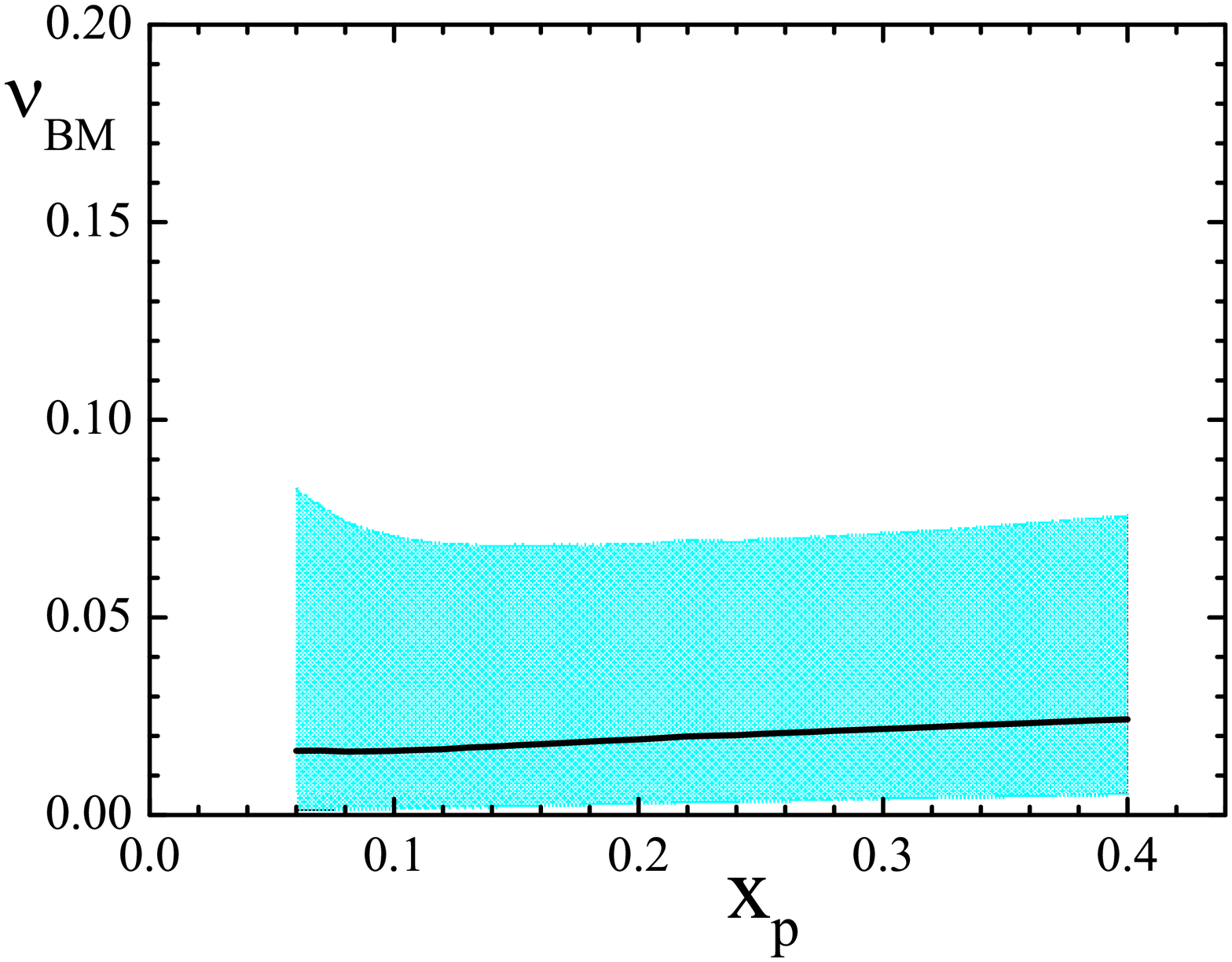}
  \includegraphics[width=0.48\columnwidth]{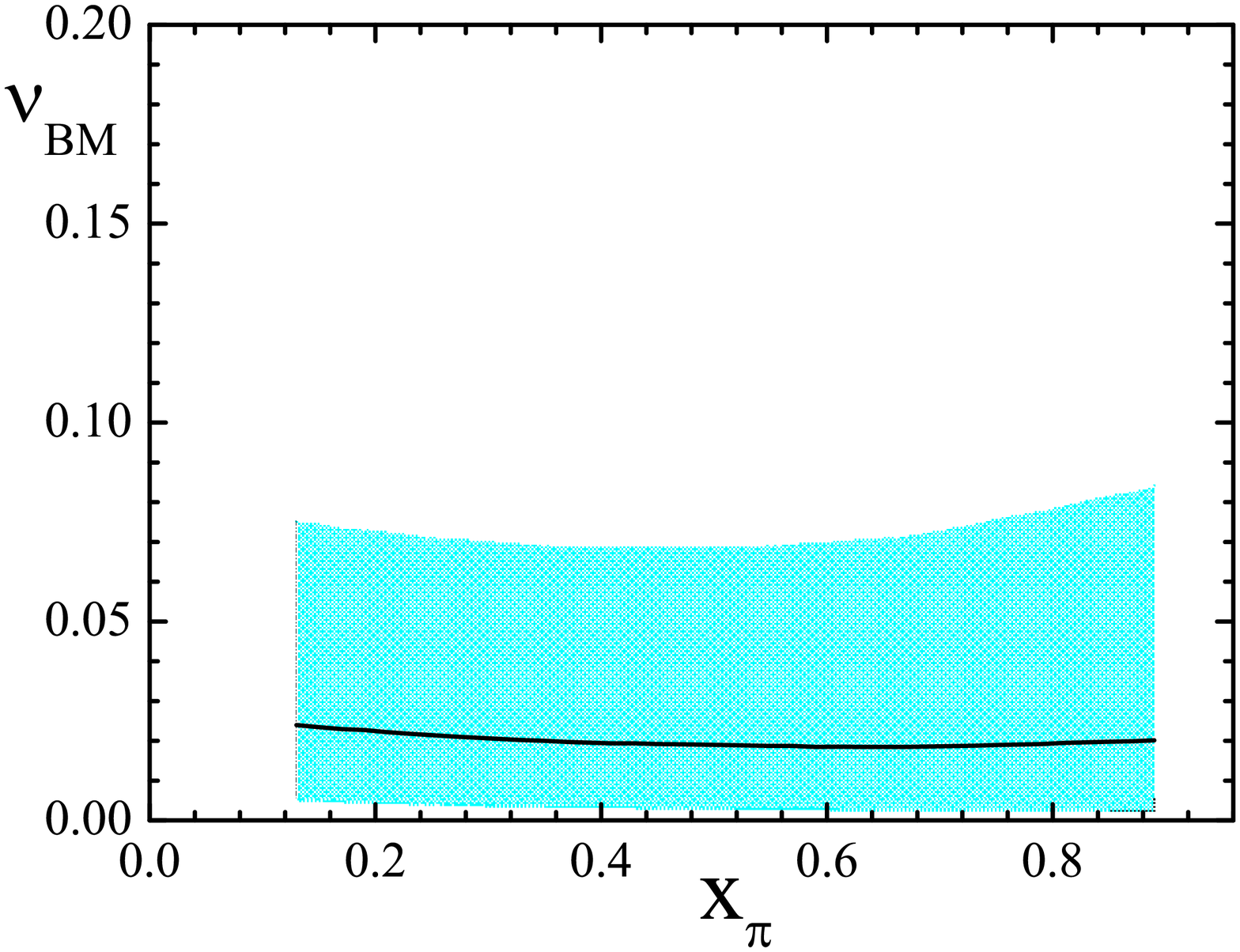}\\
  \includegraphics[width=0.48\columnwidth]{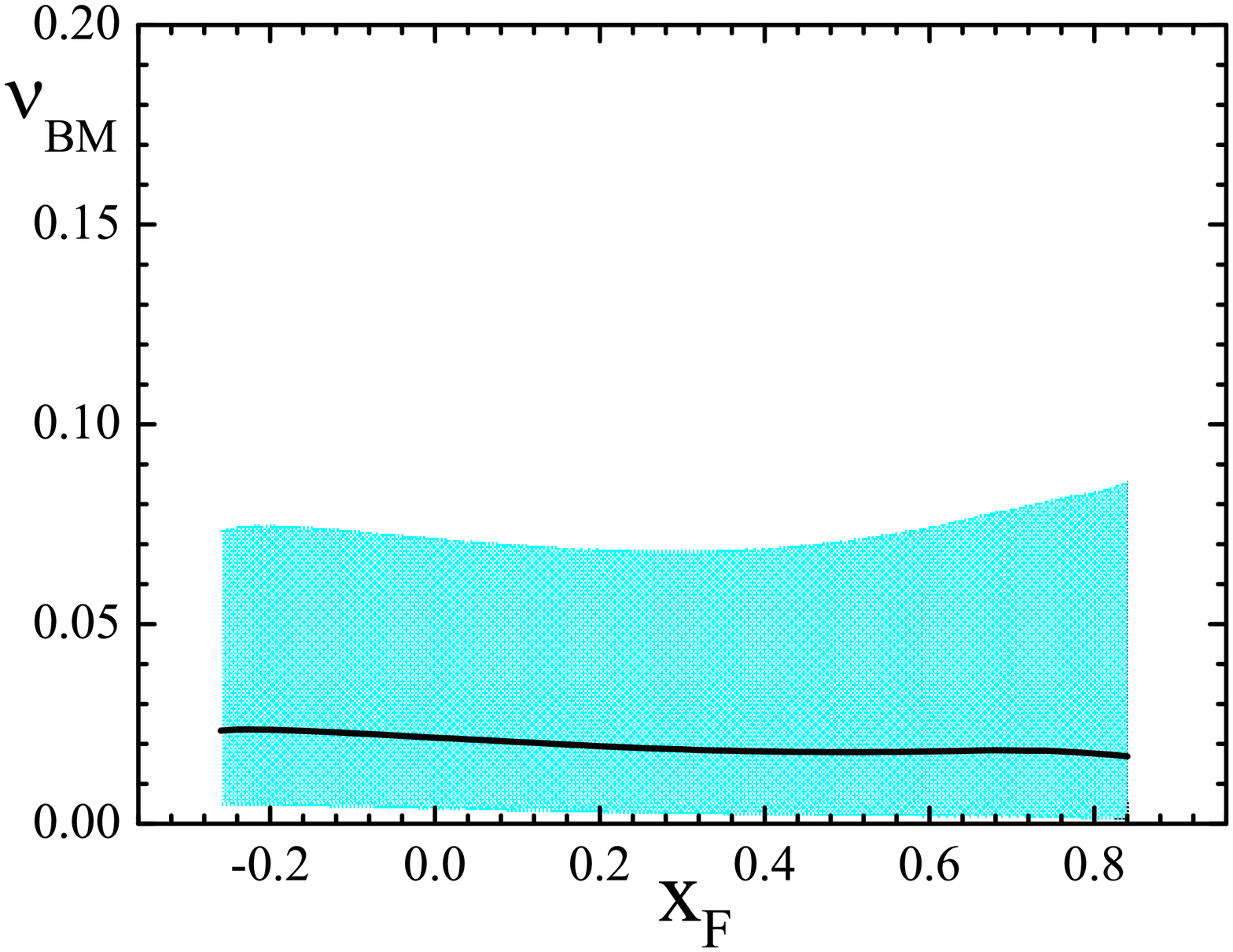}
  \includegraphics[width=0.48\columnwidth]{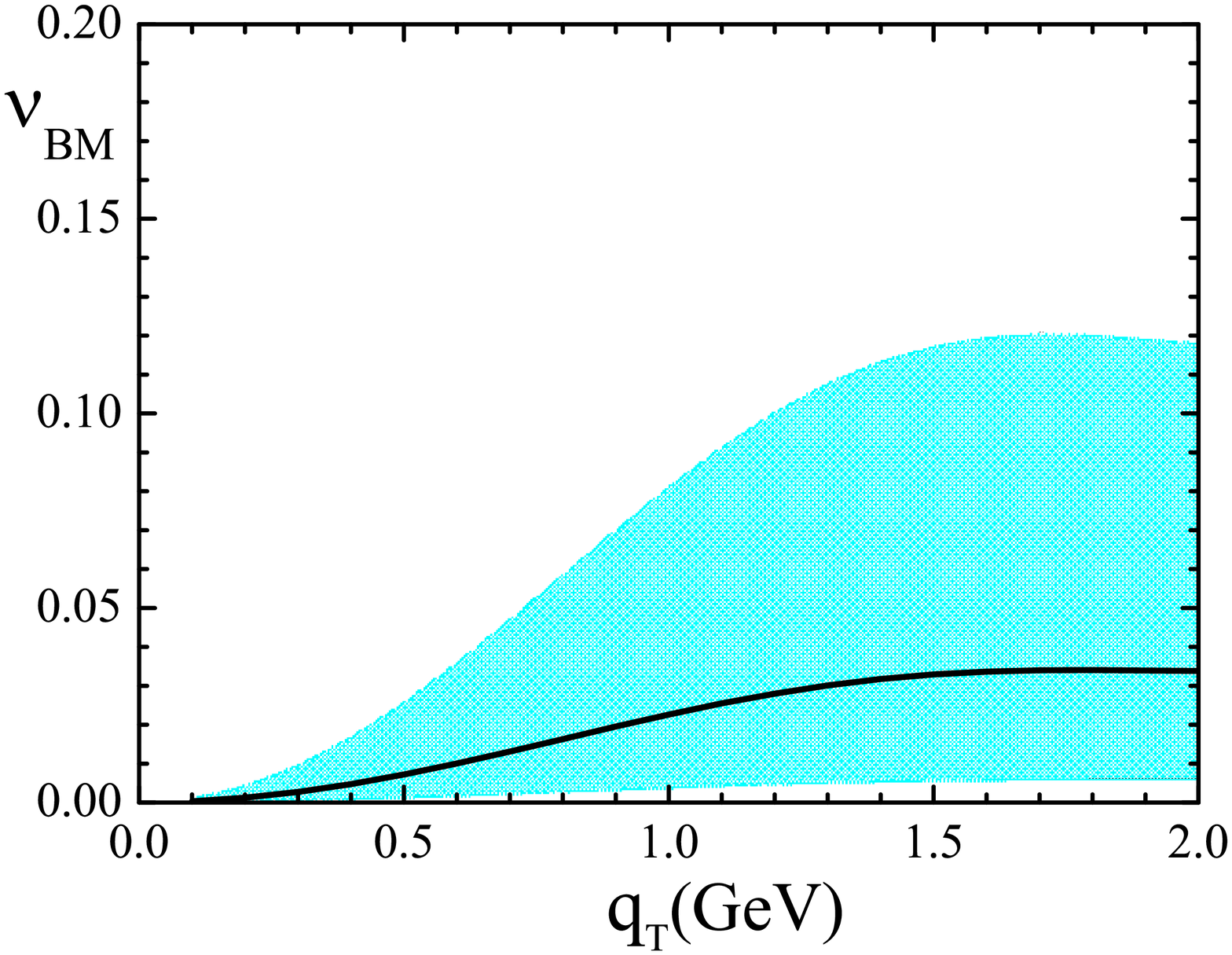}
  \caption{The $\cos2\phi$ azimuthal asymmetries $\nu_{BM}$ as the functions of $x_p$~(upper left), $x_\pi$~(upper right), $x_F$~(lower left) and $q_\perp$~(lower right) for the unpolarized $\pi p$ Drell-Yan process considering the TMD evolution in the kinematical region of COMPASS. The shadow areas correspond to the uncertainty of the parameters in the parametrization of the Boer-Mulders function for proton in Ref.~\cite{Lu:2009ip}. Figures from Ref.~\cite{Wang:2018naw}.}
  \label{fig:asy}
\end{figure}

Using Eq.~(\ref{eq:numerator}), the $\cos 2\phi$ azimuthal asymmetry contributed by the double Boer-Mulders effect in the $\pi N$ Drell-Yan process was analyzed in Ref.~\cite{Wang:2018naw}, in which the TMD evolution of the Boer-Mulders function was included.
In this calculation, the Boer-Mulders function of the proton was chosen from the parametrization in Ref.~\cite{Lu:2009ip} at the initial energy $Q_0^2=1\mathrm{GeV}^2$.
As mentioned in Sec.~\ref{sec:bm_pion}, the Boer-Mulders function of the pion wals adopted from the model calculation in Ref.~\cite{Wang:2017onm}.
Here we plot the estimated asymmetry $\nu_{BM}$ as function of $x_p,\ x_\pi,\ x_F$ and $q_\perp$ in the kinematical region of COMPASS in Fig.~\ref{fig:asy}.
The bands correspond to the uncertainty of the parametrization of the Boer-Mulders function of the proton~\cite{Lu:2009ip}.
We find from the plots that, in the TMD formalism, the $\cos2\phi$ azimuthal asymmetry in the unpolarized $\pi^-p$ Drell-Yan process contributed by the Boer-Mulders functions is around several percent.
Although the uncertainty from the proton Boer-Mulders functions is rather large, the asymmetry is firmly positive in the entire kinematical region.
The asymmetries as the functions of $x_p,\ x_\pi,\ x_F$ show slight dependence on the variables, while the $q_\perp$ dependent asymmetry shows increasing tendency along with the increasing $q_\perp$ in the small $q_\perp$ range where the TMD formalism is valid.
The result in Figs.~\ref{fig:asy} indicates that, precise measurements on the Boer-Mulders asymmetry $\nu_{BM}$ as functions of $x_p,\ x_\pi,\ x_F$ and $q_\perp$ can provide an opportunity to access the Boer-Mulders function of the pion meson.
Furthermore, the work may also shed light on the proton Boer-Mulders function since the previous extractions on it were mostly performed without TMD evolution.

\section{Summary and prospects}
\label{Sec.Conclusion}

It has been a broad consensus that the study on the TMD observables will provide information on the partons' intrinsic transverse motions inside a hadron.
In the previous sections we have tried to substantiate this statement mainly focusing on the unpolarized and single-polarized $\pi p$ Drell-Yan process within the TMD factorization. In particular, we reviewed the extraction of the nonperturbative function from the Drell-Yan and SIDIS data in the evolution formalism of the TMD distributions.
We also discussed the further applications of the TMD factorization in the phenomenology of unpolarized cross section, the Sivers asymmetry, and the $\cos2\phi$ azimuthal asymmetry in the $\pi p$ Drell-Yan process.
In summary, we have the following understanding on the $\pi N$ Drell-Yan from the viewpoint of the TMD factorization:
\begin{itemize}
\item
The extraction of nonperturbative Sudakov form factor from the $\pi N$ Drell-Yan may shed light on the evolution (scale dependence) of the pion TMD distribution.
The prediction on transverse momentum distribution of the dilepton in the small $q_\perp$ region is compatible with the COMPASS measurement, and may serve as a first step to study the spin/azimuthal asymmetry in the $\pi N$ Drell-Yan process at COMPASS.
\item
The precise measurement on the single-spin asymmetry in the kinematical region of COMPASS can provide great opportunity to access the Sivers function. Besides the TMD evolution effect, the choice of the scale dependence of the Qiu-Sterman function can affect the shape of the asymmetry and should be considered in the future extraction of the Sivers function.
\item
Sizable $\cos2\phi$ asymmetry contributed by the convolution of the Boer-Mulders functions of the pion meson and the proton can still be observed at COMPASS after the TMD evolution effect is considered.
Future data with higher accuracy may provide further constraint on the Boer-Mulders function of the pion meson as well as that of the proton.
\end{itemize}

Although a lot of progress on the theoretical framework of the TMD factorization and TMD evolution has been made, the improvement is still necessary both from the perturbative and nonperturbative aspects. In the future, the study of $S_{\rm NP}$ based on more precise experimental data is needed, such as including the flavor dependence and hadron dependence on the functional form for $S_{\rm NP}$. From the viewpoint of the perturbative region, higher-order calculation of the hard factors and coefficients will improve the accuracy of the theoretical framework. Moreover, most of the numerical calculations are based on the approximation that the $Y$-term correction is negligible in the small transverse momentum region, the inclusion of this term in the future estimate could be done to test the magnitude of the term. In addition, the TMD factorization is suitable to describe the small transverse momentum physics, while the collinear factorization is suitable for the large transverse momentum or the integrated transverse momentum. The matching between the two factorization schemes to study the unpolarized and polarized process over the whole transverse momentum region may be also necessary~\cite{Collins:2016hqq,Gamberg:2017jha}.

\section*{Acknowledgements}
This work is partially supported by the NSFC (China) grant 11575043 and by the Fundamental Research Funds for the Central Universities of China. X. Wang is supported by the NSFC (China) grant 11847217 and the China Postdoctoral Science Foundation under Grant No. 2018M640680.


\begin{thebibliography}{99}

\bibitem{Christenson:1970um}
  J.~H.~Christenson, G.~S.~Hicks, L.~M.~Lederman, P.~J.~Limon, B.~G.~Pope and E.~Zavattini,
  Phys.\ Rev.\ Lett.\  {\bf 25}, 1523 (1970).


\bibitem{Drell:1970wh}
  S.~D.~Drell and T.~M.~Yan,
  Phys.\ Rev.\ Lett.\  {\bf 25}, 316 (1970)
  Erratum: [Phys.\ Rev.\ Lett.\  {\bf 25}, 902 (1970)].

\bibitem{Bordalo:1987cs}
  P.~Bordalo {\it et al.} [NA10 Collaboration],
  Phys.\ Lett.\ B {\bf 193}, 368 (1987).

\bibitem{Betev:1985pf}
  B.~Betev {\it et al.} [NA10 Collaboration],
  Z.\ Phys.\ C {\bf 28}, 9 (1985).

\bibitem{Falciano:1986wk}
  S.~Falciano {\it et al.} [NA10 Collaboration],
  Z.\ Phys.\ C {\bf 31}, 513 (1986).

\bibitem{Guanziroli:1987rp}
  M.~Guanziroli {\it et al.} [NA10 Collaboration],
  Z.\ Phys.\ C {\bf 37}, 545 (1988).

\bibitem{Conway:1989fs}
  J.~S.~Conway {\it et al.},
  Phys.\ Rev.\ D {\bf 39}, 92 (1989).
\bibitem{Palestini:1985zc}
  S.~Palestini {\it et al.},
  Phys.\ Rev.\ Lett.\  {\bf 55}, 2649 (1985).

\bibitem{Anassontzis:1987hk}
  E.~Anassontzis {\it et al.},
  Phys.\ Rev.\ D {\bf 38}, 1377 (1988).

\bibitem{Gautheron:2010wva}
  F.~Gautheron {\it et al.} [COMPASS Collaboration],
  SPSC-P-340, CERN-SPSC-2010-014.
\bibitem{Aghasyan:2017jop}
  M.~Aghasyan {\it et al.} (COMPASS Collaboration),
  Phys.\ Rev.\ Lett.\  {\bf 119}, 112002 (2017)
  [arXiv:1704.00488 [hep-ex]].
\bibitem{Collins:1984kg}
  J.~C.~Collins, D.~E.~Soper and G.~F.~Sterman,
  Nucl.\ Phys.\ {\bf B250}, 199 (1985).

\bibitem{Ji:2004xq}
  X.~D.~Ji, J.~P.~Ma and F.~Yuan,
  Phys.\ Lett.\ B {\bf 597}, 299 (2004).

\bibitem{Ji:2004wu}
  X.~D.~Ji, J.~P.~Ma and F.~Yuan,
  Phys.\ Rev.\ D {\bf 71}, 034005 (2005)
  [hep-ph/0404183].

\bibitem{Collins:2011zzd}
J.~Collins, Foundations of perturbative QCD, Cambridge University Press, 2011.

\bibitem{Collins:2002kn}
  J.~C.~Collins,
  Phys.\ Lett.\ B {\bf 536}, 43 (2002)
  [hep-ph/0204004].

\bibitem{Ji:2002aa}
  X.~D.~Ji and F.~Yuan,
  Phys.\ Lett.\ B {\bf 543}, 66 (2002).

  \bibitem{Belitsky:2002sm}
  A.~V.~Belitsky, X.~Ji and F.~Yuan,
  Nucl.\ Phys.\ {\bf B656}, 165 (2003)
  [hep-ph/0208038].

\bibitem{Boer:2003cm}
  D.~Boer, P.~J.~Mulders and F.~Pijlman,
  Nucl.\ Phys.\ {\bf B667}, 201 (2003)
  [hep-ph/0303034].

\bibitem{Mulders:1995dh}
  P.~J.~Mulders and R.~D.~Tangerman,
  Nucl.\ Phys.\ {\bf B461}, 197 (1996)
  Erratum: [Nucl.\ Phys.\ {\bf B484}, 538 (1997)]
  [hep-ph/9510301].

\bibitem{Bacchetta:2006tn}
  A.~Bacchetta, M.~Diehl, K.~Goeke, A.~Metz, P.~J.~Mulders and M.~Schlegel,
  J. High Energy Phys. 02 (2007) 093
  [hep-ph/0611265].

  \bibitem{Collins:1981uk}
  J.~C.~Collins and D.~E.~Soper,
  Nucl.\ Phys.\ {\bf B193}, 381 (1981),
   Erratum: [Nucl.\ Phys.\ {\bf B213}, 545 (1983)].

 \bibitem{Aybat:2011zv}
  S.~M.~Aybat and T.~C.~Rogers,
  Phys.\ Rev.\ D {\bf 83}, 114042 (2011)
  [arXiv:1101.5057 [hep-ph]].
\bibitem{Collins:1999dz}
  J.~C.~Collins and F.~Hautmann,
  Phys.\ Lett.\ B {\bf 472}, 129 (2000).
\bibitem{Collins:2012uy}
  J.~C.~Collins and T.~C.~Rogers,
  Phys.\ Rev.\ D {\bf 87}, 034018 (2013).

 \bibitem{Echevarria:2012pw}
  M.~G.~Echevarria, A.~Idilbi, A.~Sch\"{a}fer and I.~Scimemi,
  Eur.\ Phys.\ J.\ C {\bf 73}, 2636 (2013)
  [arXiv:1208.1281 [hep-ph]].
\bibitem{Pitonyak:2013dsu}
  D.~Pitonyak, M.~Schlegel and A.~Metz,
  Phys.\ Rev.\ D {\bf 89}, 054032 (2014).
\bibitem{Boer:2008fr}
  D.~Boer,
  Nucl.\ Phys.\ {\bf B806}, 23 (2009).

\bibitem{Arnold:2008kf}
  S.~Arnold, A.~Metz and M.~Schlegel,
  Phys.\ Rev.\ D {\bf 79}, 034005 (2009).

  \bibitem{Lambertsen:2016wgj}
  M.~Lambertsen and W.~Vogelsang,
  Phys.\ Rev.\ D {\bf 93}, 114013 (2016)
  [arXiv:1605.02625 [hep-ph]].

\bibitem{Ji:2006ub}
  X.~Ji, J.~W.~Qiu, W.~Vogelsang and F.~Yuan,
  Phys.\ Rev.\ Lett.\  {\bf 97}, 082002 (2006)
  [hep-ph/0602239].

\bibitem{Ji:2006vf}
  X.~Ji, J.~W.~Qiu, W.~Vogelsang and F.~Yuan,
  Phys.\ Rev.\ D {\bf 73}, 094017 (2006)
  [hep-ph/0604023].

\bibitem{Sivers:1989cc}
  D.~W.~Sivers,
  Phys.\ Rev.\ D {\bf 41}, 83 (1990).

\bibitem{Brodsky:2002cx}
  S.~J.~Brodsky, D.~S.~Hwang and I.~Schmidt,
  Phys.\ Lett.\ B {\bf 530}, 99 (2002)
  [hep-ph/0201296].

\bibitem{Brodsky:2002rv}
  S.~J.~Brodsky, D.~S.~Hwang and I.~Schmidt,
  Nucl.\ Phys.\ {\bf B642}, 344 (2002)
  [hep-ph/0206259].

\bibitem{Anselmino:2009st}
  M.~Anselmino, M.~Boglione, U.~D'Alesio, S.~Melis, F.~Murgia and A.~Prokudin,
  Phys.\ Rev.\ D {\bf 79}, 054010 (2009)
  [arXiv:0901.3078 [hep-ph]].

\bibitem{Kang:2009bp}
  Z.~B.~Kang and J.~W.~Qiu,
  Phys.\ Rev.\ Lett.\  {\bf 103}, 172001 (2009)
  [arXiv:0903.3629 [hep-ph]].

\bibitem{Peng:2014hta}
  J.~C.~Peng and J.~W.~Qiu,
  Prog.\ Part.\ Nucl.\ Phys.\  {\bf 76}, 43 (2014)
  [arXiv:1401.0934 [hep-ph]].

 \bibitem{Echevarria:2014xaa}
  M.~G.~Echevarria, A.~Idilbi, Z.~B.~Kang and I.~Vitev,
  Phys.\ Rev.\ D {\bf 89}, 074013 (2014)
  [arXiv:1401.5078 [hep-ph]].
\bibitem{Huang:2015vpy}
  J.~Huang, Z.~B.~Kang, I.~Vitev and H.~Xing,
  Phys.\ Rev.\ D {\bf 93}, 014036 (2016)
  [arXiv:1511.06764 [hep-ph]].

\bibitem{Anselmino:2016uie}
  M.~Anselmino, M.~Boglione, U.~D'Alesio, F.~Murgia and A.~Prokudin,
  J. High Energy Phys. 04 (2017) 046
  [arXiv:1612.06413 [hep-ph]].

\bibitem{Fermilab1} L.D.~Isenhower {\it et al.},\\
\url{http://www.fnal.gov/directorate/program_planning/June2012Public/P-1027_Pol-Drell-Yan-proposal.pdf}.

\bibitem{Fermilab2}
   D.~Geesaman{\it et al.},\\
  \url{http://www.fnal.gov/directorate/program_planning/June2013PACPublic/P-1039_LOI_polarized_DY.pdf}.
\bibitem{ANDY}
    E.C.~Aschenauer{\it et al.},
   \url{https://www.bnl.gov/npp/docs/pac0611/DY_pro_110516_final.2.pdf}.
\bibitem{Adamczyk:2015gyk}
  L.~Adamczyk {\it et al.} (STAR Collaboration),
  Phys.\ Rev.\ Lett.\  {\bf 116}, 132301 (2016)
  [arXiv:1511.06003 [nucl-ex]].
\bibitem{Adolph:2016dvl}
  C.~Adolph {\it et al.} (COMPASS Collaboration),
  Phys.\ Lett.\ B {\bf 770}, 138 (2017)
  [arXiv:1609.07374 [hep-ex]].

\bibitem{Lam:1978pu}
  C.~S.~Lam and W.~K.~Tung,
Phys.\ Rev.\ D {\bf 18}, 2447 (1978).

\bibitem{Aaltonen:2011nr}
  T.~Aaltonen {\it et al.} [CDF Collaboration],
  Phys.\ Rev.\ Lett.\  {\bf 106}, 241801 (2011)
  [arXiv:1103.5699 [hep-ex]].

\bibitem{Khachatryan:2015paa}
  V.~Khachatryan {\it et al.} [CMS Collaboration],
  Phys.\ Lett.\ B {\bf 750}, 154 (2015)
  [arXiv:1504.03512 [hep-ex]].

\bibitem{bran93}
A. Brandenburg, O. Nachtmann and E. Mirkes, Z.~Phys.~C {\bf 60}, 697 (1993).

\bibitem{bran94}
A. Brandenburg, S.J. Brodsky, V.V. Khoze and D. M\"uller, Phys.~Rev.~Lett.~{\bf 73},
939 (1994).

\bibitem{Eskola}
K.J. Eskola, P. Hoyer, M. V\"antinnen and R. Vogt,
Phys.~Lett.~B {\bf 333}, 526 (1994);\\
J.G. Heinrich {\em et al.}, Phys.~Rev.~D {\bf 44}, 1909 (1991).

\bibitem{Peng:2015spa}
  J.~C.~Peng, W.~C.~Chang, R.~E.~McClellan and O.~Teryaev,
  Phys.\ Lett.\ B {\bf 758}, 384 (2016)
  [arXiv:1511.08932 [hep-ph]].

  \bibitem{Boer:2006eq}
  D.~Boer and W.~Vogelsang,
  Phys.\ Rev.\ D {\bf 74}, 014004 (2006).
\bibitem{Chang:2018pvk}
  W.~C.~Chang, R.~E.~McClellan, J.~C.~Peng and O.~Teryaev,
  arXiv:1811.03256 [hep-ph].

\bibitem{Boer:1999mm}
  D.~Boer,
  Phys.\ Rev.\ D {\bf 60}, 014012 (1999)
  [hep-ph/9902255].

\bibitem{Boer:1997nt}
  D.~Boer and P.~J.~Mulders,
  Phys.\ Rev.\ D {\bf 57}, 5780 (1998)
  [hep-ph/9711485].
\bibitem{Idilbi:2004vb}
  A.~Idilbi, X.~d.~Ji, J.~P.~Ma and F.~Yuan,
  Phys.\ Rev.\ D {\bf 70}, 074021 (2004)
  [hep-ph/0406302].

\bibitem{Collins:2014jpa}
  J.~Collins and T.~Rogers,
  Phys.\ Rev.\ D {\bf 91}, 074020 (2015)
  [arXiv:1412.3820 [hep-ph]].

\bibitem{Collins:2016hqq}
  J.~Collins, L.~Gamberg, A.~Prokudin, T.~C.~Rogers, N.~Sato and B.~Wang,
  Phys.\ Rev.\ D {\bf 94}, 034014 (2016)
  [arXiv:1605.00671 [hep-ph]].

  \bibitem{Bacchetta:2017gcc}
  A.~Bacchetta, F.~Delcarro, C.~Pisano, M.~Radici and A.~Signori,
  J. High Energy Phys. 06 (2017) 081
  [arXiv:1703.10157 [hep-ph]].

\bibitem{Bacchetta:2013pqa}
  A.~Bacchetta and A.~Prokudin,
  Nucl.\ Phys.\ {\bf B875}, 536 (2013)
  [arXiv:1303.2129 [hep-ph]].
 \bibitem{Collins:1981uw}
  J.~C.~Collins and D.~E.~Soper,
  Nucl.\ Phys.\ {\bf B194}, 445 (1982).
 \bibitem{Kang:2011mr}
  Z.~B.~Kang, B.~W.~Xiao and F.~Yuan,
  Phys.\ Rev.\ Lett.\  {\bf 107}, 152002 (2011)
  [arXiv:1106.0266 [hep-ph]].
 \bibitem{Aybat:2011ge}
  S.~M.~Aybat, J.~C.~Collins, J.~W.~Qiu and T.~C.~Rogers,
  Phys.\ Rev.\ D {\bf 85}, 034043 (2012)
  [arXiv:1110.6428 [hep-ph]].
 \bibitem{Echevarria:2014rua}
  M.~G.~Echevarria, A.~Idilbi and I.~Scimemi,
  Phys.\ Rev.\ D {\bf 90}, 014003 (2014)
  [arXiv:1402.0869 [hep-ph]].
 \bibitem{Landry:2002ix}
  F.~Landry, R.~Brock, P.~M.~Nadolsky and C.~P.~Yuan,
  Phys.\ Rev.\ D {\bf 67}, 073016 (2003)
  [hep-ph/0212159].
\bibitem{Qiu:2000ga}
  J.~W.~Qiu and X.~F.~Zhang,
  Phys.\ Rev.\ Lett.\  {\bf 86}, 2724 (2001)
  [hep-ph/0012058].
 \bibitem{Korchemsky:1994is}
  G.~P.~Korchemsky and G.~F.~Sterman,
  Nucl.\ Phys.\ {\bf B437}, 415 (1995)
  [hep-ph/9411211].

 \bibitem{Kang:2015msa}
  Z.~B.~Kang, A.~Prokudin,
  P.~Sun and F.~Yuan,
  Phys.\ Rev.\ D {\bf 93}, 014009 (2016).

 \bibitem{Sun:2013dya}
  P.~Sun and F.~Yuan,
  Phys.\ Rev.\ D {\bf 88}, 034016 (2013)
  [arXiv:1304.5037 [hep-ph]].


\bibitem{Su:2014wpa}
  P.~Sun, J.~Isaacson, C.-P.~Yuan and F.~Yuan,
  Int.\ J.\ Mod.\ Phys.\ A {\bf 33}, no. 11, 1841006 (2018)
  [arXiv:1406.3073 [hep-ph]].

\bibitem{Konychev:2005iy}
  A.~V.~Konychev and P.~M.~Nadolsky,
  Phys.\ Lett.\ B {\bf 633}, 710 (2006)
  [hep-ph/0506225].

  \bibitem{Davies:1984sp}
  C.~T.~H.~Davies, B.~R.~Webber and W.~J.~Stirling,
  Nucl.\ Phys.\ {\bf B256}, 413 (1985).

\bibitem{Ellis:1997sc}
  R.~K.~Ellis, D.~A.~Ross and S.~Veseli,
  Nucl.\ Phys.\ {\bf B503}, 309 (1997)
  [hep-ph/9704239].

\bibitem{Anselmino:2012aa}
  M.~Anselmino, M.~Boglione and S.~Melis,
  Phys.\ Rev.\ D {\bf 86}, 014028 (2012)
  [arXiv:1204.1239 [hep-ph]].

\bibitem{Aidala:2014hva}
  C.~A.~Aidala, B.~Field, L.~P.~Gamberg and T.~C.~Rogers,
  Phys.\ Rev.\ D {\bf 89}, 094002 (2014)
  [arXiv:1401.2654 [hep-ph]].


\bibitem{Collins:1985xx}
  J.~C.~Collins and D.~E.~Soper,
  Nucl.\ Phys.\  {\bf B284}, 253 (1987).

\bibitem{Wang:2017zym}
  X.~Wang, Z.~Lu and I.~Schmidt,
  J. High Energy Phys. 08 (2017) 137
  [arXiv:1707.05207 [hep-ph]].

\bibitem{Stirling:1993gc}
  W.~J.~Stirling and M.~R.~Whalley,
  J.\ Phys.\ G {\bf 19}, D1 (1993).

\bibitem{James:1975dr}
  F.~James and M.~Roos,
  Comput.\ Phys.\ Commun.\  {\bf 10}, 343 (1975).

\bibitem{James:1994vla}
  F.~James,
  CERN-D-506.

\bibitem{Ceccopieri:2018nop}
  F.~A.~Ceccopieri, A.~Courtoy, S.~Noguera and S.~Scopetta,
  Eur.\ Phys.\ J.\ C {\bf 78}, 644 (2018)
  [arXiv:1801.07682 [hep-ph]].

  \bibitem{Sun:2013hua}
  P.~Sun and F.~Yuan,
  Phys.\ Rev.\ D {\bf 88}, 114012 (2013)
  [arXiv:1308.5003 [hep-ph]].

  \bibitem{Qiu:1991pp}
  J.~W.~Qiu and G.~F.~Sterman,
  Phys.\ Rev.\ Lett.\  {\bf 67}, 2264 (1991).
  \bibitem{Qiu:1991wg}
  J.~W.~Qiu and G.~F.~Sterman,
  Nucl.\ Phys.\ {\bf B378}, 52 (1992).
\bibitem{Qiu:1998ia}
  J.~W.~Qiu and G.~F.~Sterman,
  Phys.\ Rev.\ D {\bf 59}, 014004 (1999)
  [hep-ph/9806356].
\bibitem{Anselmino:2012re}
  M.~Anselmino, M.~Boglione and S.~Melis,
  arXiv:1209.1541 [hep-ph].

\bibitem{Boglione:2018dqd}
  M.~Boglione, U.~D'Alesio, C.~Flore and J.~O.~Gonzalez-Hernandez,
  J. High Energy Phys. 07 (2018) 148
  [arXiv:1806.10645 [hep-ph]].


\bibitem{Zhang:2008nu}
  B.~Zhang, Z.~Lu, B.~Q.~Ma and I.~Schmidt,
  Phys.\ Rev.\ D {\bf 77}, 054011 (2008)
  [arXiv:0803.1692 [hep-ph]].


\bibitem{Lu:2009ip}
  Z.~Lu and I.~Schmidt,
  Phys.\ Rev.\ D {\bf 81}, 034023 (2010).

\bibitem{Barone:2009hw}
  V.~Barone, S.~Melis and A.~Prokudin,
  Phys.\ Rev.\ D {\bf 81}, 114026 (2010)
  [arXiv:0912.5194 [hep-ph]].

\bibitem{Barone:2010gk}
  V.~Barone, S.~Melis and A.~Prokudin,
  Phys.\ Rev.\ D {\bf 82}, 114025 (2010)
  [arXiv:1009.3423 [hep-ph]].

  \bibitem{Prokudin:2015ysa}
  A.~Prokudin, P.~Sun and F.~Yuan,
  Phys.\ Lett.\ B {\bf 750}, 533 (2015)
  [arXiv:1505.05588 [hep-ph]].

\bibitem{Collins:2004nx}
  J.~C.~Collins and A.~Metz,
  Phys.\ Rev.\ Lett.\  {\bf 93}, 252001 (2004).
\bibitem{Mantry:2009qz}
  S.~Mantry and F.~Petriello,
  Phys.\ Rev.\ D {\bf 81}, 093007 (2010).

\bibitem{Becher:2010tm}
  T.~Becher and M.~Neubert,
  Eur.\ Phys.\ J.\ C {\bf 71}, 1665 (2011).

\bibitem{GarciaEchevarria:2011rb}
  M.~G.~Echevarria, A.~Idilbi and I.~Scimemi,
  J. High Energy Phys. 07 (2012) 002.

  \bibitem{Chiu:2012ir}
  J.~Y.~Chiu, A.~Jain, D.~Neill and I.~Z.~Rothstein,
  J. High Energy Phys. 05 (2012) 084.





\bibitem{Ji:2014hxa}
  X.~Ji, P.~Sun, X.~Xiong and F.~Yuan,
  Phys.\ Rev.\ D {\bf 91}, 074009 (2015)
  [arXiv:1405.7640 [hep-ph]].

\bibitem{Catani:2000vq}
  S.~Catani, D.~de Florian and M.~Grazzini,
  Nucl.\ Phys.\ {\bf B596}, 299 (2001)
  [hep-ph/0008184].

\bibitem{Nadolsky:1999kb}
  P.~M.~Nadolsky, D.~R.~Stump and C.~P.~Yuan,
  Phys.\ Rev.\ D {\bf 61}, 014003 (2000)
  Erratum: [Phys.\ Rev.\ D {\bf 64}, 059903 (2001)]  [hep-ph/9906280].
\bibitem{Koike:2006fn}
  Y.~Koike, J.~Nagashima and W.~Vogelsang,
  Nucl.\ Phys.\ {\bf B744}, 59 (2006)
  [hep-ph/0602188].
\bibitem{Collins:1977iv}
  J.~C.~Collins and D.~E.~Soper,
  Phys.\ Rev.\ D {\bf 16}, 2219 (1977).

\bibitem{Zhu:2006gx}
  L.~Y.~Zhu {\it et al.} (NuSea Collaboration),
  Phys.\ Rev.\ Lett.\  {\bf 99}, 082301 (2007)
  [hep-ex/0609005].

  \bibitem{Zhu:2008sj}
  L.~Y.~Zhu {\it et al.} (NuSea Collaboration),
  Phys.\ Rev.\ Lett.\  {\bf 102}, 182001 (2009)
  [arXiv:0811.4589 [nucl-ex]].

\bibitem{Collins:1978yt}
  J.~C.~Collins,
  Phys.\ Rev.\ Lett.\  {\bf 42}, 291 (1979).

\bibitem{Chiappetta:1986yg}
P.~Chiappetta and M.~Le~Bellac,
\newblock Z. Phys. C {\bf32}, 521 (1986).

\bibitem{Blazek}
M. Blazek, M. Biyajima and N. Suzuki, Z.~Phys.~C {\bf43}, 447 (1989).

\bibitem{Zhou:2009rp}
  J.~Zhou, F.~Yuan and Z.~T.~Liang,
  Phys.\ Lett.\  B {\bf 678}, 264 (2009)
  [arXiv:0901.3601 [hep-ph]].

\bibitem{Lu:2016pdp}
  Z.~Lu,
  Front.\ Phys.\ (Beijing) {\bf 11}, no. 1, 111204 (2016).

\bibitem{Sutton:1991ay}
  P.~J.~Sutton, A.~D.~Martin, R.~G.~Roberts and W.~J.~Stirling,
  Phys.\ Rev.\ D {\bf 45}, 2349 (1992).

\bibitem{Wang:2018naw}
  X.~Wang, W.~Mao and Z.~Lu,
  Eur.\ Phys.\ J.\ C {\bf 78}, 643 (2018)
  [arXiv:1805.03017 [hep-ph]].

\bibitem{Wang:2018pmx}
  X.~Wang and Z.~Lu,
  Phys.\ Rev.\ D {\bf 97}, 054005 (2018)
  [arXiv:1801.00660 [hep-ph]].
\bibitem{Kang:2012em}
  Z.~B.~Kang and J.~W.~Qiu,
  Phys.\ Lett.\ B {\bf 713}, 273 (2012)
  [arXiv:1205.1019 [hep-ph]].

\bibitem{Kang:2008ey}
  Z.~B.~Kang and J.~W.~Qiu,
  Phys.\ Rev.\ D {\bf 79}, 016003 (2009)
  [arXiv:0811.3101 [hep-ph]].

\bibitem{Zhou:2008mz}
  J.~Zhou, F.~Yuan and Z.~T.~Liang,
  Phys.\ Rev.\ D {\bf 79}, 114022 (2009)
  [arXiv:0812.4484 [hep-ph]].

\bibitem{Vogelsang:2009pj}
  W.~Vogelsang and F.~Yuan,
  Phys.\ Rev.\ D {\bf 79}, 094010 (2009)
  [arXiv:0904.0410 [hep-ph]].

\bibitem{Braun:2009mi}
  V.~M.~Braun, A.~N.~Manashov and B.~Pirnay,
  Phys.\ Rev.\ D {\bf 80}, 114002 (2009)
   Erratum: [Phys.\ Rev.\ D {\bf 86}, 119902 (2012)]
  [arXiv:0909.3410 [hep-ph]].

\bibitem{Ma:2011nd}
  J.~P.~Ma and H.~Z.~Sang,
  J. High Energy Phys. 04 (2011) 062
  [arXiv:1102.2679 [hep-ph]].


\bibitem{Schafer:2012ra}
  A.~Schafer and J.~Zhou,
  Phys.\ Rev.\ D {\bf 85}, 117501 (2012)
  [arXiv:1203.5293 [hep-ph]].


\bibitem{Ma:2012xn}
  J.~P.~Ma and Q.~Wang,
  Phys.\ Lett.\ B {\bf 715}, 157 (2012)
  [arXiv:1205.0611 [hep-ph]].


\bibitem{Zhou:2015lxa}
  J.~Zhou,
  Phys.\ Rev.\ D {\bf 92},  074016 (2015)
  [arXiv:1507.02819 [hep-ph]].

\bibitem{Salam:2008qg}
  G.~P.~Salam and J.~Rojo,
  Comput.\ Phys.\ Commun.\  {\bf 180}, 120 (2009)
  [arXiv:0804.3755 [hep-ph]].

\bibitem{Wang:2017onm}
  Z.~Wang, X.~Wang and Z.~Lu,
  Phys.\ Rev.\ D {\bf 95}, 094004 (2017)
  [arXiv:1702.03637 [hep-ph]].

\bibitem{Botje:2010ay}
  M.~Botje,
  Comput.\ Phys.\ Commun.\  {\bf 182}, 490 (2011)
  [arXiv:1005.1481 [hep-ph]].

\bibitem{Airapetian:2009ae}
  A.~Airapetian {\it et al.}  [HERMES Collaboration],
  Phys.\ Rev.\ Lett.\  {\bf 103}, 152002 (2009)
  [arXiv:0906.3918 [hep-ex]].

\bibitem{Alekseev:2008aa}
  M.~Alekseev {\it et al.}  [COMPASS Collaboration],
  Phys.\ Lett.\ B {\bf 673}, 127 (2009)
  [arXiv:0802.2160 [hep-ex]].

\bibitem{Adolph:2012sp}
  C.~Adolph {\it et al.}  [COMPASS Collaboration],
  Phys.\ Lett.\ B {\bf 717}, 383 (2012)
  [arXiv:1205.5122 [hep-ex]].

\bibitem{Qian:2011py}
  X.~Qian {\it et al.}  [Jefferson Lab Hall A Collaboration],
  Phys.\ Rev.\ Lett.\  {\bf 107}, 072003 (2011)
  [arXiv:1106.0363 [nucl-ex]].
\bibitem{Aschenauer:2013woa}
  E.~C.~Aschenauer {\it et al.},
  arXiv:1304.0079 [nucl-ex].
  \bibitem{Gamberg:2017jha}
  L.~Gamberg, A.~Metz, D.~Pitonyak and A.~Prokudin,
  Phys.\ Lett.\ B {\bf 781}, 443 (2018).

\end{thebibliography}
\end{document}